\documentclass[twocolumn]{aastex6}

\usepackage{graphics}
\usepackage{epsfig}
\usepackage{graphicx}
\usepackage{color}
\usepackage{pstricks}
\usepackage{afterpage}
\usepackage{epstopdf}
\usepackage{natbib}

\newcommand{\unit}[1]{\ifmmode\,{\rm #1}\else$\,{\rm #1}$\fi}

\newcommand{\etal}{~{et~al.}\ }  
\newcommand{\msun}{M_{\odot}}

\def\kms{km~s$^{-1}$}
\def\arcmin{$^{\prime}$}
\def\arcsec{$^{\prime\prime}$}
\def\msun{$M_\odot$}
\def\mhi{$M_{HI}$}

\newcommand{\hi}{H{\sc\,i}}

\usepackage{natbib}

\usepackage{longtable}

\received{2016 October 29}
\revised{2017 March 15}
\accepted{2017 May 25}

\shorttitle{\hi\ Bearing UDGs}
\shortauthors{Leisman\etal}

\begin{document}

\title{(Almost) Dark Galaxies in the ALFALFA Survey: Isolated \hi\ Bearing Ultra Diffuse Galaxies}
\author{Lukas Leisman\altaffilmark{1},
Martha P. Haynes\altaffilmark{1},
Steven Janowiecki\altaffilmark{2},
Gregory Hallenbeck\altaffilmark{3},
Gyula J\'{o}zsa\altaffilmark{4,5,6},
Riccardo Giovanelli\altaffilmark{1},
Elizabeth A. K. Adams\altaffilmark{7},
David Bernal Neira\altaffilmark{8},
John M. Cannon\altaffilmark{9},
William F. Janesh\altaffilmark{10},
Katherine L. Rhode\altaffilmark{10},
John J. Salzer\altaffilmark{10}
}
\altaffiltext{1}{Cornell Center for Astrophysics and Planetary Science, 
Space Sciences Building, Cornell University, Ithaca, NY 14853, USA}
\altaffiltext{2}{International Centre for Radio Astronomy Research
(ICRAR), University of Western Australia, 35 Stirling Highway, Crawley,
WA 6009, Australia}
\altaffiltext{3}{Department of Physics and Astronomy, Union College,
  Schenectady, NY 12308, USA}
\altaffiltext{4}{SKA South Africa Radio Astronomy Research Group, 3rd Floor, The Park, Park Road, Pinelands 7405, South Africa}
\altaffiltext{5}{Rhodes Centre for Radio Astronomy Techniques \& Technologies, Department of Physics and Electronics, Rhodes University, PO Box 94, Grahamstown 6140, South Africa}
\altaffiltext{6}{ArgelanderInstitut f\"ur Astronomie, Auf dem H\"ugel 71, D-53121 Bonn, Germany}
\altaffiltext{7}{ASTRON, the Netherlands Institute for Radio Astronomy, Postbus 2, 7990 AA, Dwingeloo, The Netherlands}
\altaffiltext{8}{Departamento de F\'isica, Universidad de los Andes, Cra. 1 No. 18A-10, Edificio Ip, Bogot\'a, Colombia}
\altaffiltext{9}{Department of Physics \& Astronomy, Macalester College, 1600 Grand Avenue, Saint Paul, MN 55105}
\altaffiltext{10}{Department of Astronomy, Indiana University, 727 East
 Third Street, Bloomington, IN 47405, USA}

\begin{abstract}
We present a sample of 115 very low optical surface brightness, highly extended, \hi-rich galaxies carefully selected from the ALFALFA survey that have similar optical absolute magnitudes, surface brightnesses, and radii to recently discovered ``ultra-diffuse" galaxies (UDGs). However, these systems are bluer and have more irregular morphologies than other UDGs, are isolated, and contain significant reservoirs of \hi.  We find that while these sources have normal star formation rates for \hi\ selected galaxies of similar stellar mass, they have very low star formation efficiencies. 
We further present deep optical and \hi\ synthesis follow up imaging of three of these \hi-bearing ultra-diffuse sources. We measure \hi\ diameters extending to $\sim$40~kpc, but note that while all three sources have large \hi\ diameters for their stellar mass, they are consistent with the \hi\ mass - \hi\ radius relation.
We further analyze the \hi\ velocity widths and rotation velocities for the unresolved and resolved sources respectively, and find that the sources appear to inhabit halos of dwarf galaxies. We estimate spin parameters, and suggest that these sources may exist in high spin parameter halos, and as such may be potential \hi-rich progenitors to the ultra-diffuse galaxies observed in cluster environments.
\end{abstract}


\section{Introduction}
\label{intro}

Recent advances in low optical surface brightness survey techniques (e.g., \citealp{abraham14a}) have unveiled substantial populations of very low surface brightness ``ultra-diffuse" galaxies (UDGs),
which have stellar masses of dwarfs ($\lesssim$10$^8$\msun), but sizes comparable to L$_{\star}$ galaxies (effective radii of several kpc; \citealp{vandokkum15a}). 

UDGs appear to be common in cluster environments (e.g., \citealp{koda15a}; \citealp{vanderburg16a}), and have colors and morphologies consistent with extrapolation of 
early type galaxies on the red sequence \citep{vandokkum15a}. 
But UDGs also appear to exist outside of clusters.  \cite{martinez-delgado16a} report the discovery of a UDG in the Pisces-Perseus Filament, and \cite{merritt16a} and \cite{castelli16a} report the discovery of UDGs in group environments. \cite{roman16a} statistically estimate the distribution of UDGs around Abell~168, and suggest that more than 50\% of UDGs could exist outside of the cluster environment. 

UDGs appear to have high dark matter fractions within their optical radii, but the distribution of their halo masses is still unclear. \cite{vandokkum15a}  suggest UDGs could be failed L$_{\star}$ galaxies, with star formation quenched early in their lifetime, and \cite{vandokkum16a}  use stellar spectroscopy to estimate the halo mass of the UDG Dragonfly 44 to be near that of the Milky Way ($\sim$10$^{12}$\msun). However, \cite{beasley16a} used spectroscopy of globular clusters, and  \cite{peng16a}, \cite{beasley16b}, and \cite{amorisco16b} use globular cluster counts to suggest instead that UDGs are more likely to reside in dwarf halos similar to the Large Magellanic Cloud ($\lesssim$10$^{11}$\msun). \cite{zaritsky16a} uses scaling relations to suggest that it is also possible that UDGs span a range of halo masses between these extremes. 

Several hypotheses have been suggested to explain these enigmatic galaxies. Some mechanisms focus on environmental effects. \cite{yozin15a} demonstrate that they can reproduce the properties of UDGs in simulations where UDGs are satellites of clusters, falling into the cluster early, around z$\sim$2, and \cite{baushev16a} and \cite{burkert16a} invoke 2-body tidal encounters in dense environments. Other explanations suggest that UDGs formed via internal mechanisms.
\cite{amorisco16a} suggest they likely represent sources in halos in the high end tail of the spin parameter distribution, and \cite{dicintio16a} reproduce the extended stellar distributions of UDGs in isolated dwarf halos using gas outflows.

These latter explanations predict that UDGs could potentially exist in isolated environments, 
 contain large reservoirs of gas, and be actively forming stars. \cite{dicintio16a} explicitly predict non-negligible \hi\ gas masses of 10$^{7-9}$\msun, and that the gas plays an important role in creating large radii. But the \hi\ contents of UDGs are uncertain; the best \hi\ upper limits at the distances of most UDGs ($\sim$100~Mpc) are around 10$^9$\msun\ (\citealp{haynes11a}; \citealp{martinez-delgado16a}).

Further, if there are isolated star forming UDGs, they may be difficult to recognize. UDGs are a subset of ``classical" low surface brightness galaxies (e.g., \citealp{schombert92a}), which are known to exist across a wide range of sizes (e.g. \citealp{zucker06a}; \citealp{bothun87a}) and environments (e.g. \citealp{impey88a}; \citealp{impey96a}), and range from star forming late type galaxies \citep{mcgaugh95a} to bulge dominated early types \citep{beijersbergen99a}. While classical LSB galaxies are typically higher surface brightness or less extended than UDGs, \cite{yagi16a} point out that a small number of these LSB sources fit the observationally defined selection criteria for UDGs, a few of which are late type and contain \hi. However, they suggest that they must be rare in the field due to the small number of detected sources.

Yet, finding isolated low surface brightness ultra-diffuse sources optically in a systematic way is difficult due to the lack of easily attainable distance information, and often relies on color selection criteria. Still, these sources may be detectable at other wavelengths if they contain significant gas. 

The largest volume blind \hi\ survey to date, the Arecibo Legacy Fast ALFA  (Arecibo L-band Feed Array) extragalactic \hi\ survey (e.g., \citealp{giovanelli05a}; \citealp{haynes11a}) is well-positioned to locate low surface brightness sources missed by optical detection algorithms \citep{du15a}. 
Here we explore isolated ultra-diffuse sources from the ALFALFA survey which match the optical selection criteria for previously reported UDGs, and present results on three ultra-diffuse ALFALFA sources that happened to be included in exploratory observations by 
the ALFALFA (Almost) Darks campaign (e.g. \citealp{cannon15a}). This campaign has been exploring the 1\% of sources not easily identified with optical counterparts in the Sloan Digital Sky Survey (SDSS) or Digitized Sky Survey 2 (DSS2).
We note that these ``(almost) dark" observations have already uncovered at least one ultra-diffuse source with a similarly large radius for its stellar mass. \cite{janowiecki15a} report the detection of AGC~229385, which has a peak g-band surface brightness of 26.5~mag~asec$^{-2}$ and a half light radius of $\sim$2.4~kpc (assuming a distance of 25~Mpc). This source appears even more diffuse than most other reported UDGs, though it also has a significant distance uncertainty. 

The paper is outlined as follows: we describe the selection of \hi-bearing UDGs 
from the overall ALFALFA population in section \ref{sample} and our data in section \ref{data}. We then present optical and \hi\ results in section \ref{results}. We discuss the star formation and dark matter halos of these sources in section \ref{discussion} and conclude in section \ref{conclusions}. For all calculations, the assumed cosmology is $H_0 = 70~\unit{km s^{-1} Mpc^{-1}}$,
$\Omega_m=0.3$, and $\Omega_\Lambda=0.7$.

\section{Sample Selection}
\label{sample}
\begin{figure*}
\centering
\includegraphics[width=\textwidth]{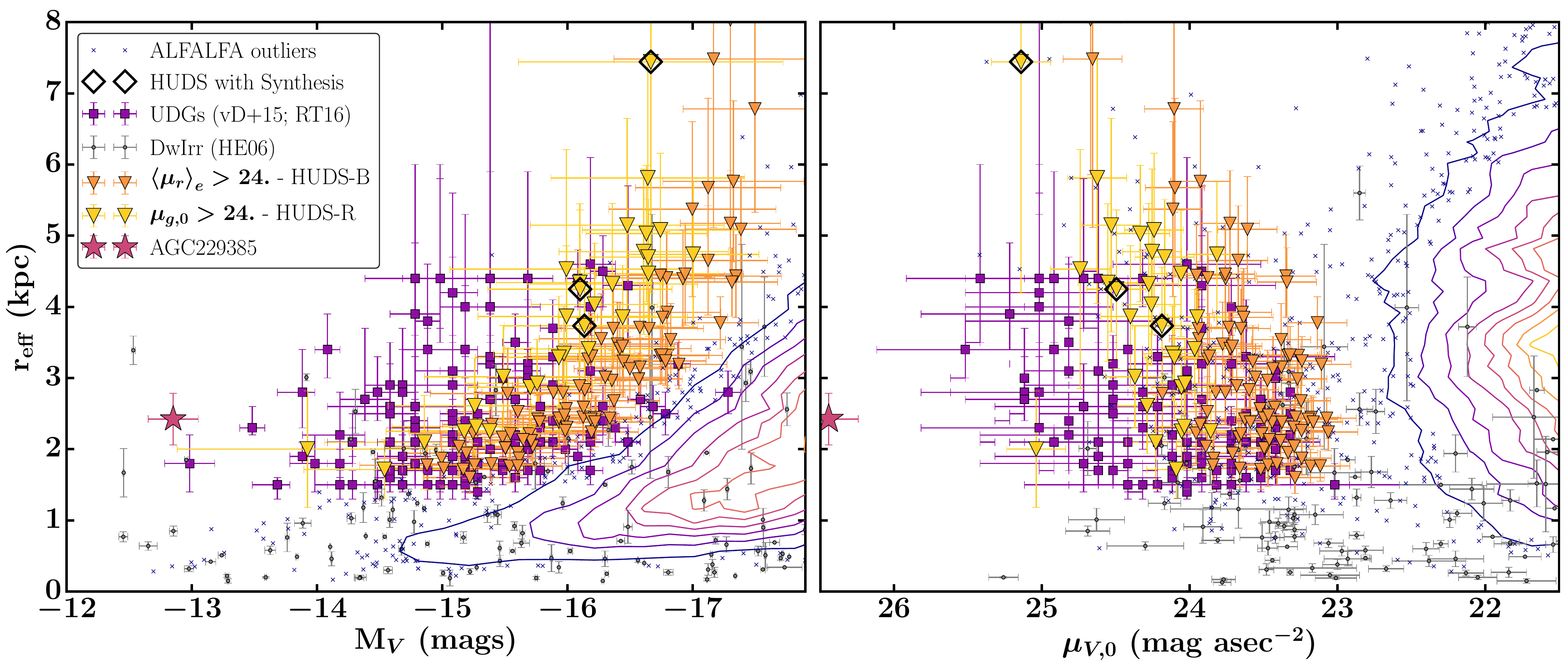}\\
\caption{Optical properties of \hi-bearing ultra-diffuse ALFALFA sources (HUDS) in comparison with other ``ultra-diffuse" galaxies, showing that they fall in a similar part of parameter space to other UDGs. HUDS conforming to the stricter definition of ``ultra-diffuse" (HUDS-R) are shown as lighter yellow triangles; those satisfying the broader criteria (HUDS-B) are shown with darker orange triangles. HUDS with existing synthesis observations are marked with black diamonds. Comparison samples of dwarf irregulars are small grey points (\citealp{hunter06a}), and UDGs are purple squares (\citealp{vandokkum15a}; \citealp{roman16a}). Other ALFALFA sources are shown by small dark blue points, and contours increasing in 10\% intervals. 
\label{compare}
}
\end{figure*}

%
There are 24,159 high signal to noise, clearly extragalactic sources in the ALFALFA 70\% catalog, 22,940 of which fall within the SDSS footprint and are at least 10\arcmin\ away from stars in the Yale Bright Star Catalog. We use this sample to search
for \hi-bearing, isolated, ultra-diffuse galaxies as described below.

\subsection{Distance and Isolation Selection Criteria}
\label{sample.iso}
Due to Arecibo's comparatively large beam size (3.5\arcmin), cross identification with optical surveys becomes more difficult at larger distances. We thus restrict our search for UDGs to sources within 120~Mpc, where the ALFALFA beam corresponds to $\sim$120~kpc, or about 3$\times$ the diameter of the detected sources discussed below. This distance cut is also important to maximize physical resolution for future follow up observations. We also set a minimum distance limit of 25~Mpc, since redshift-dependent distance estimates for sources closer than 25~Mpc are subject to significant uncertainty.
 
Most optically dark or (almost) dark \hi\ features turn out to be tidal in origin. To eliminate potential confusion between satellites and central halos, we restrict our sample to isolated sources by requiring that the nearest neighbor within 500~\kms\ in the Arecibo General Catalog\footnotemark[1] has a projected separation of at least 350~kpc. This eliminates potential confusion with low surface brightness tidal dwarf galaxies (e.g., \citealp{lee-waddell16a}), and extended tidal debris (e.g., \citealp{leisman16a}). These distance and isolation criteria reduce our potential sample to 5186 sources.

\footnotetext[1]{The Arecibo General Catalog is a private database maintained over the years by MPH and RG; within the ALFALFA volume it contains all bright galaxies and galaxies of known redshift as available in NED with cz $<$ 18000~\kms\ (including all measurements from SDSS and ALFALFA), and additional unpublished \hi\ results as they are acquired.}

\subsection{Optical Selection Criteria}
\label{sample.opt}

\begin{figure*}[t!]
\centering
{\bf \large \hspace{0.57in} DF 17 \hspace{1.22in} AGC~122966 \hspace{0.98in} AGC~334315}\\
\includegraphics[width=0.3\textwidth]{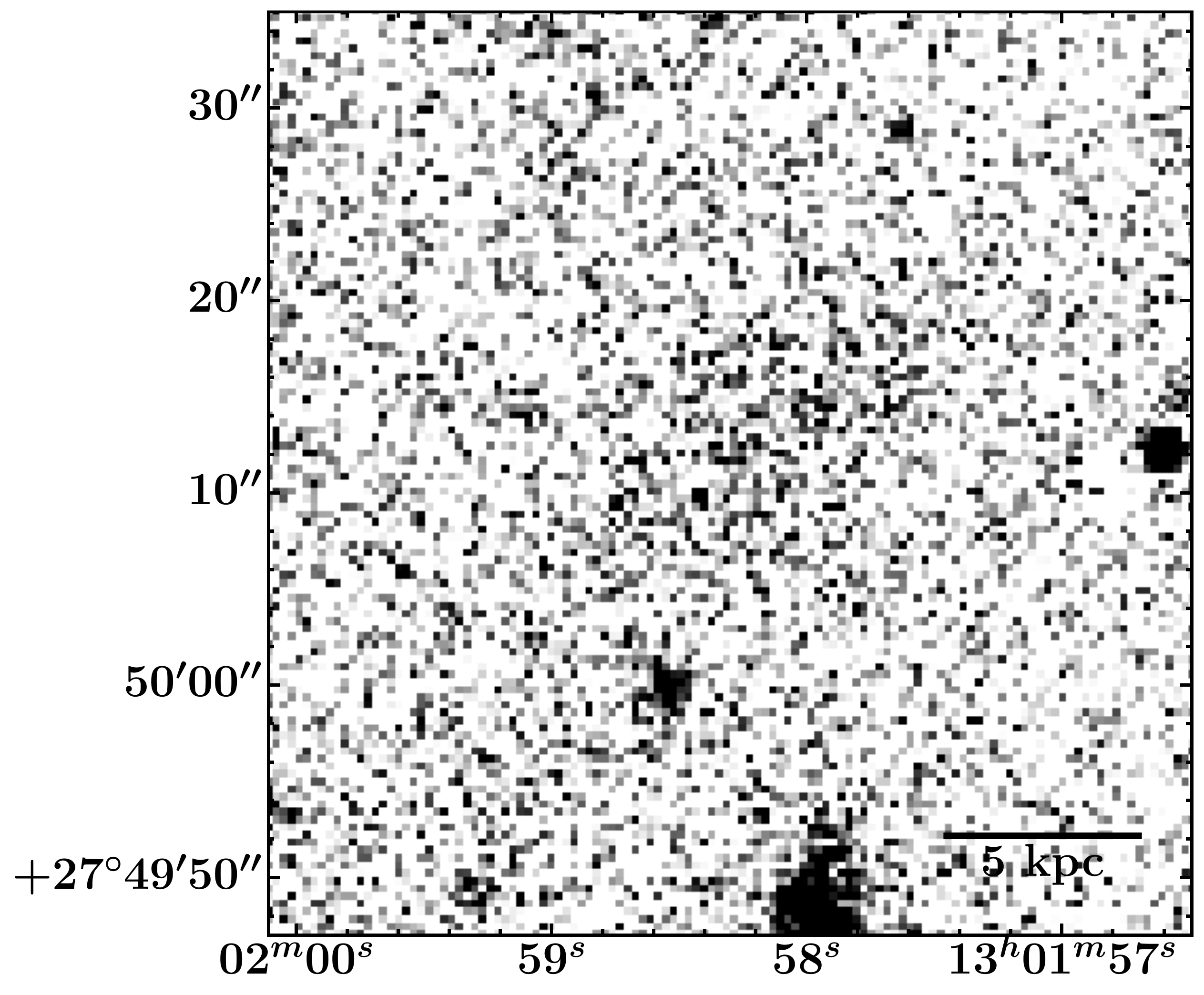}
\includegraphics[width=0.3\textwidth]{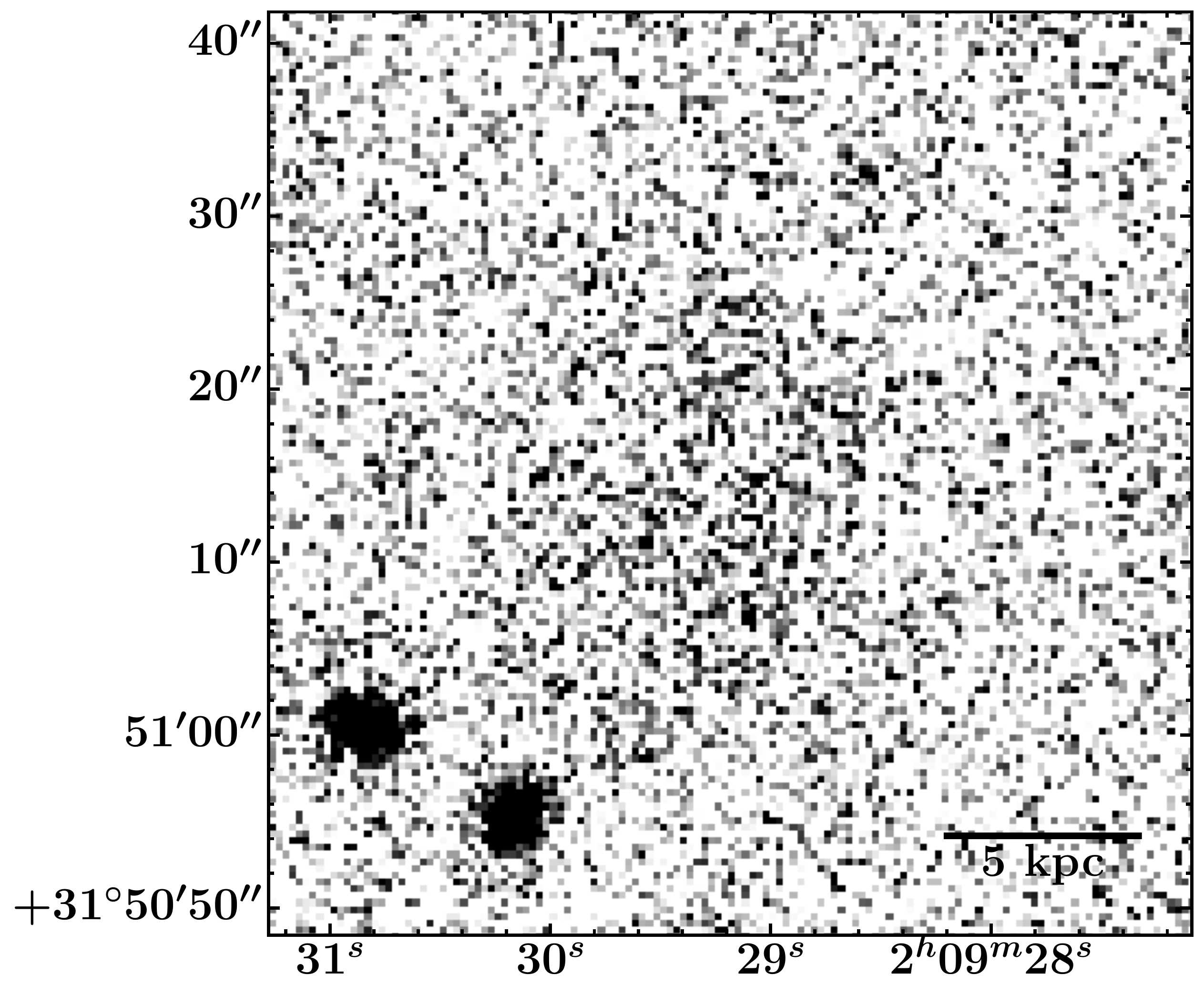}
\includegraphics[width=0.3\textwidth]{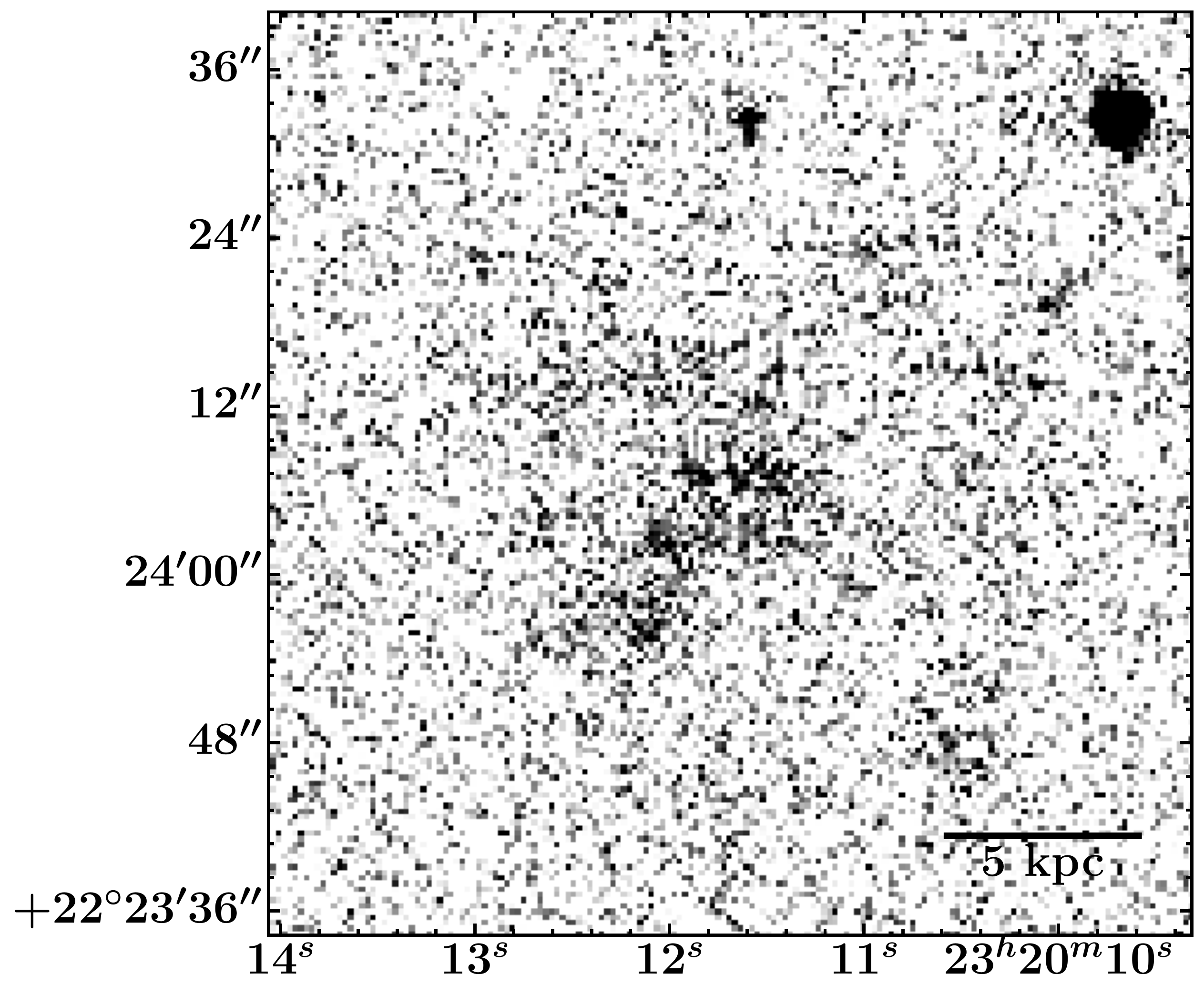}\\
\includegraphics[width=0.3\textwidth]{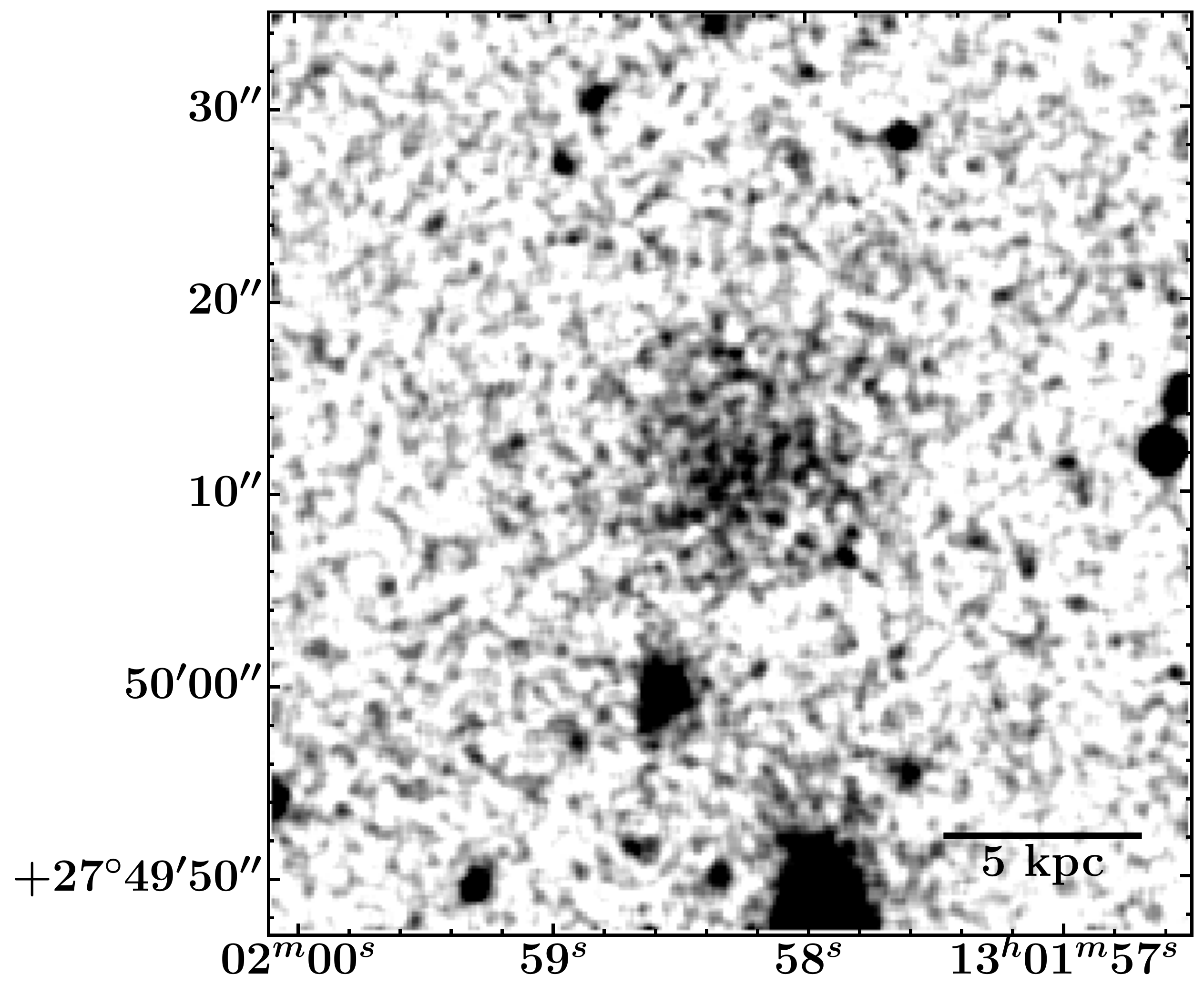}
\includegraphics[width=0.3\textwidth]{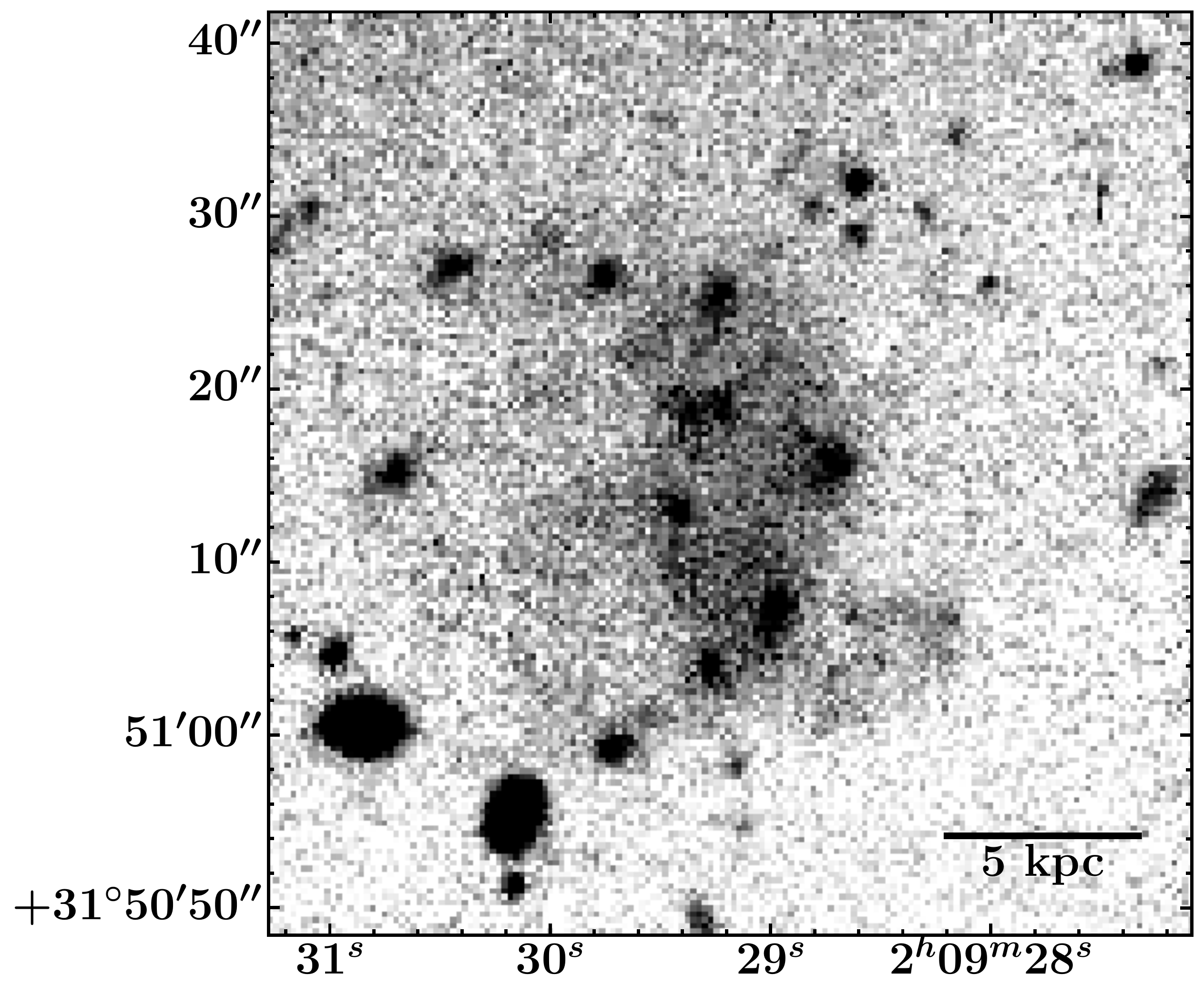}
\includegraphics[width=0.3\textwidth]{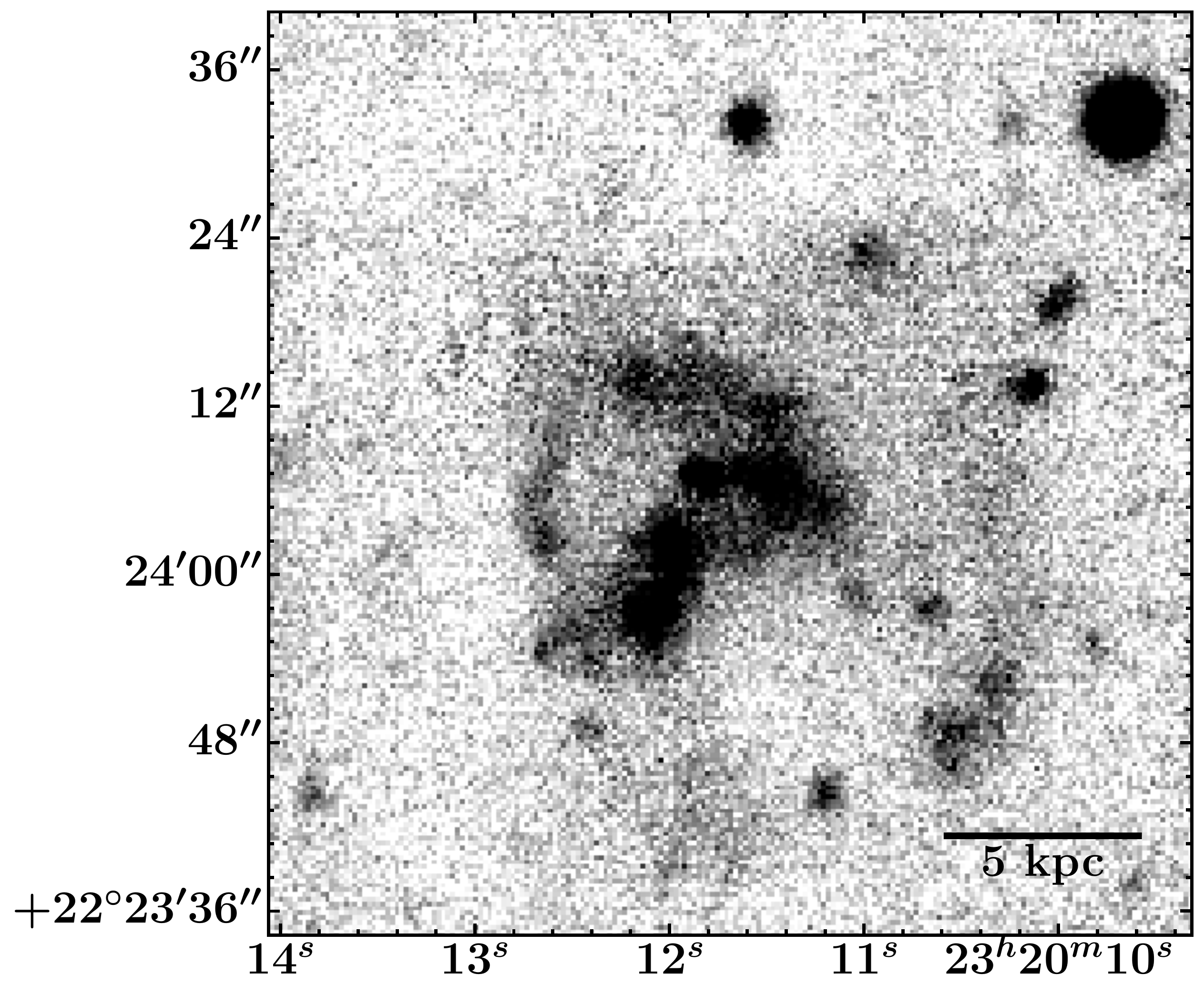}
\vspace{-0.1cm}
\caption{Comparison of SDSS and deeper imaging for a UDG and two HUDS; these sources are only barely detected in SDSS, but are located and confirmed by the position prior from ALFALFA. 
Left: SDSS (top) and CFHT (bottom) imaging of the Coma cluster UDG DF17. The UDG is visible in SDSS, and remains smooth in deeper imaging. Center and Right: SDSS (top) and WIYN pODI (bottom) imaging of two HUDS, AGC~122966 and 334315. The HUDS have a similar surface brightness to the Coma UDG, but have significantly more complicated morphologies in deeper optical imaging. RA and Dec are in J2000 coordinates.
\label{SDSSodiCompare}
}
\end{figure*}

Sources that fit the criteria for ultra-diffuse galaxies are barely detected at the depth of the SDSS data, and thus tend to have poor or missing measurements in the SDSS catalog. However, automated measurements from the SDSS catalog tend to be reasonably reliable for moderate to high surface brightness galaxies. Thus, we use a two step selection process to find ultra-diffuse sources.

First, we eliminate moderate or high surface brightness sources from our sample with matching SDSS DR12 catalog measurements in the most sensitive filters (g, r, and i bands). 
Specifically, we eliminate sources that have an average surface brightness within the measured exponential effective radius 
$<$23.8~mag~arcsec$^{-2}$ and an average petrosian surface brightness within the 90\% petrosian radius 
$<$25.0~mag~arcsec$^{-2}$ in all 3 bands.
This cut leaves 645 candidate sources that do not meet all 6 criteria. We visually inspect these sources to remove clear high surface brightness sources with bad catalog measurements, and sources with missing or bad SDSS data (due to, e.g., bright stars), leaving  $\sim$200 low surface brightness candidates.

Second, we perform our own photometry on SDSS images  of the remaining sources, correcting for galactic extinction and the effects of the PSF, but not for cosmological dimming (see section \ref{data.opt.sdss}). We use these measurements to select  sources with similar absolute magnitude, surface brightnesses, and radii to previously reported UDGs. 
The definition of ``ultra-diffuse" varies significantly in the literature. \cite{vandokkum15a} originally define their sample as having central, g-band surface brightness $\mu_{g,0}\gtrsim$24~mags~arcsec$^{-2}$, and 1.5$<r_{\rm eff}<$4.6~kpc. %
However, other authors have explored a wider range of parameter space (see \citealp{yagi16a} for a useful summary). For example, \cite{vanderburg16a} use the average r-band surface brightness enclosed within the effective radius,  24.0 $ \le \left< \mu(r,r_{\rm eff}) \right> \le 26.5$ mag arcsec$^{-2}$ (note: $\left< \mu(r,r_{\rm eff}) \right>$ is 1.12 mag~arcsec$^{-2}$ brighter than $\mu_{r,0}$  for an exponential profile, though for the average UDG g-r color of $\sim$0.5, this approximately corresponds to $\mu_{g,0}\gtrsim$23.4~mag~arcsec$^{-2}$). Some authors have also suggested restrictions in absolute magnitude, luminosity, or stellar mass (e.g., \citealp{mihos15a}), explicitly limiting UDGs to dwarf mass stellar populations. Differences in color and profile shape further complicate the matter, since
sources detected in \hi\ are usually star forming, with bluer colors and clumpier morphologies than previously reported UDGs. 
Thus, we choose to define a more restrictive and less restrictive sample, but note that our choice of what constitutes ``ultra-diffuse" is somewhat arbitrary. 

 Specifically, we select a restrictive sample of 30 \hi-bearing ultra-diffuse sources (HUDS-R), with half light radii r$_{g,{\rm eff}}>$1.5~kpc, $\mu_{g,0}>24$~mag~arcsec$^{-2}$, and M$_g>-$16.8~mag, and a broader sample (HUDS-B) of 115 sources with r$_{r,{\rm eff}}>1.5$~kpc, $\left< \mu(r,r_{\rm eff}) \right> >24$~mag~arcsec$^{-2}$, and M$_r>-$17.6 (corresponding to the surface brightness and radius limits from \citealp{vandokkum15a} and \citealp{vanderburg16a} respectively; since these papers do not give explicit absolute magnitude limits, we chose the restrictive and broad samples to include absolute magnitudes up to that of the Small Magellanic Cloud (SMC; see \citealp{mcconnachie12a}) and 2$\times$ the SMC respectively, which are reasonable matches to other limits from the literature - see, e.g., \citealp{yagi16a}). We note that while most authors fit Sersic profiles with $n$ free, due to the low S/N of SDSS at these surface brightnesses, we have forced our fits to have exponential ($n$=1) profiles, in keeping with the average value found for UDGs 
 and typical \hi-rich galaxies.\footnotemark[1] 
We also define HUDS-BG to be the 30 HUDS-B sources that have GALEX UV observations and fall in the 40\% ALFALFA survey analyzed by \cite{huang12b}. We discuss this sample further in section \ref{data.uv}.

\footnotetext[1]{Note: some authors (e.g., \citealp{roman16a}) have suggested that a sersic index $<$1 is more appropriate for UDGs - we find that fixing n to, e.g., n=0.7 does not improve our fits, so we elect to use n=1.0.}

Figure \ref{compare} illustrates the optical similarity of these samples to other reported UDGs, and their extreme nature relative to other dwarfs and isolated sources from the ALFALFA survey. The HUDS-R and HUDS-B samples (shown with light yellow and darker orange triangles respectively), occupy a similar portion of the plot to previously reported UDGs from \cite{vandokkum15a} and \cite{roman16a} (dark purple squares). Other ALFALFA sources matching the distance and isolation criteria applied to the HUDS are shown with contours and small dark blue crosses. Dwarf irregulars from \cite{hunter06a} are shown as small grey points, emphasizing the large extent of these sources relative to typical dwarfs. 
The HUDS for which we have existing synthesis observations (section \ref{data}) are marked with black diamonds. 
We note that all sources in the figure not observed in V-band have been transformed to V-band using the Lupton filter transformations from the SDSS website\footnotemark[1].

\footnotetext[1]{\url{http://www.sdss.org/dr12/algorithms/sdssUBVRITransform/}}

We emphasize that the sources selected here differ in important ways from, e.g., the population detected in Coma by \cite{vandokkum15a}. Most importantly, the isolation criteria restrict our sample to central halos. Thus, while some UDGs may be satellite galaxies or galaxies formed via tidal interactions, this paper focuses on UDGs that are sufficiently isolated to be incompatible with these hypotheses. Further, as discussed in section \ref{results}, these HUDS tend to differ in both color and morphology from other reported UDGs. Thus, the HUDS discussed here are a specific subset of a growing population of extreme low surface brightness, ``ultra diffuse" sources.

\subsection{Sufficiency and Limitations of SDSS for Source Selection}
\label{sample.limits}
Figure \ref{compare} also illustrates the limitations of using SDSS for optical measurements. The HUDS tend to fall toward the 
brighter side of the UDG distribution, which may be surprising given that the sources were identified by their \hi\ content. Some of this bias is due to differences in the colors of the samples (discussed in section \ref{results.optical}), since their bluer relative colors increase their V-band magnitude relative to the quiescent cluster UDGs. Much of it also may be
due to the fact that HUDS are near the surface brightness limit of SDSS (which is somewhat variable, but, e.g., \cite{trujillo16a} estimate $\left< \mu_{r} \right> \ge \sim$26.5 mag~arcsec$^{-2}$ in a 10\arcsec\ $\times$ 10\arcsec\ region). Any sources with extended emission below the SDSS detection threshold but with central surface brightness just above it are likely to have their radii underestimated, and thus would be eliminated by the radius requirement. 
 Indeed, several UDG candidates are sources without easily identified counterparts observed as part of the ALFALFA (almost) darks campaign. 

However, the prior positional information from ALFALFA makes identification of UDGs in SDSS possible; sources not clearly visible in the SDSS finding chart images are in fact detected at reasonable significance in downloaded (and sometimes smoothed) images. 
The top panels of Figure \ref{SDSSodiCompare} show SDSS imaging of the UDG DF17 from \cite{vandokkum15a}, and of two HUDS from ALFALFA, shown at high contrast to emphasize low surface brightness details. The bottom panels show deep CFHT imaging of DF17, and deep pODI imaging with the WIYN 3.5m telescope of the two ALFALFA HUDS (see section \ref{data.opt.podi}). 
Of the 47 sources reported in \cite{vandokkum15a}, 46 are detected in downloaded SDSS imaging. 

Further, though the estimated parameters from SDSS data have large uncertainties, they appear sufficient for our purposes. Applying our fitting procedure to SDSS data of the 46 detected sources from \cite{vandokkum15a} produces values consistent with their measurements within estimated errors (the rms offset in central surface brightness (effective radius) is 0.4~mag (1.5~kpc), which is less than the quadrature combined average error of 0.7~mag (2.0~kpc)). Further, the SDSS results are consistent with the results of deeper imaging in the two available cases (see section \ref{data.opt.podi} and appendix A). 

We emphasize that while the SDSS data demonstrate that these sources are very low surface brightness and very extended, they
are too low signal-to-noise for detailed structural analysis, and that individual measurements are highly uncertain. Thus, this sample should only be thought of in a statistical sense. Indeed, a shift of 1$\sigma$ would move an additional 30 sources into or out of the HUDS-B sample. 

\begin{table*}[]
\caption{\textbf{Properties of HUDS}}
\vskip 5pt
\centering
\footnotesize{
\begin{tabular}{ccccccccccccc}
\hline
\hline
AGC ID & OC RA & OC Dec & c$z$    & W$_{50}$ & $\int SdV$ & Dist$^a$ & log(M$_{\rm HI}$) &    $\mu_{g,0}$      & $r_e$ & M$_g$ & g-r & Sample$^b$\\
              &  J2000 &  J2000  &  \kms\  & \kms\       &  Jy-\kms\    &Mpc &  log M$_{\odot}$  & mags/\arcsec$^2$ & kpc & mag   & mag  &    \\
      (1)    &  (2)       & (3)         &   (4)       &     (5)         &    (6)              &   (7) &            (8)                    &              (9)                & (10) & (11) & (12)  & (13) \\
\hline
322019 & 344.6121 & 1.8497 & 4819 & 33$\pm$14 & 0.54$\pm$0.07 & 72 & 8.81$\pm$0.08 & 24.62$\pm$0.18& 3.9$\pm$1.0 & -15.8$\pm$0.7 & 0.36$\pm$0.25 & R\\
103796 & 5.1650 & 6.9658 & 5647 & 31$\pm$7 & 0.48$\pm$0.04 & 80 & 8.86$\pm$0.06 & 24.23$\pm$0.14& 3.9$\pm$0.7 & -16.2$\pm$0.5 & 0.47$\pm$0.21 & R\\
113790 & 18.2587 & 27.6369 & 4952 & 31$\pm$10 & 0.33$\pm$0.03 & 69 & 8.57$\pm$0.07 & 24.31$\pm$0.14& 2.9$\pm$0.6 & -15.4$\pm$0.5 & 0.42$\pm$0.19 & R\\
114905 & 21.3271 & 7.3603 & 5435 & 27$\pm$3 & 0.96$\pm$0.04 & 76 & 9.11$\pm$0.06 & 24.85$\pm$0.19& 5.1$\pm$1.5 & -16.2$\pm$0.8 & 0.53$\pm$0.25 & R\\
114943 & 26.7775 & 7.3311 & 8416 & 32$\pm$7 & 0.40$\pm$0.04 & 116 & 9.10$\pm$0.06 & 24.48$\pm$0.18& 4.8$\pm$1.2 & -16.4$\pm$0.7 & 0.36$\pm$0.28 & R\\
113949 & 27.4108 & 30.6808 & 7380 & 44$\pm$7 & 0.44$\pm$0.05 & 102 & 9.03$\pm$0.07 & 24.29$\pm$0.18& 3.4$\pm$0.8 & -15.8$\pm$0.7 & 0.53$\pm$0.28 & R\\
122966 & 32.3708 & 31.8528 & 6518 & 35$\pm$6 & 0.53$\pm$0.04 & 90 & 9.00$\pm$0.06 & 25.37$\pm$0.23& 7.4$\pm$3.3 & -16.4$\pm$1.1 & 0.38$\pm$0.41 & R\\
\hline
\hline
\end{tabular}\label{maintable}
\begin{flushleft}
Table \ref{maintable} is published in its entirety in the machine-readable format. A portion is shown here for guidance regarding its form and content. \hi\ and optical table parameters come from ALFALFA and SDSS, and are described in sections \ref{data.hi.alfalfa} and \ref{data.opt.sdss} of the text respectively. $^a$: Distances from the ALFALFA flow model (see \citealp{haynes11a}); distance depended quantities include an error of 5~Mpc due to peculiar velocities $^b$: R = HUGS-R, B= HUGS-B
\end{flushleft}
}
\end{table*}

\section{Observations and Data}
\label{data}

All sources discussed here have available SDSS and ALFALFA data. However, 
three sources which meet the above criteria, AGC~122966, 334315, and 219533, were included in the ALFALFA (Almost) Darks campaign (e.g., \citealp{cannon15a}), and thus have deep optical and \hi\ synthesis imaging. Here we discuss the details of these observations. 

\subsection{HI Data}
\label{data.hi}

\subsubsection{ALFALFA Data}
\label{data.hi.alfalfa}
The ALFALFA observations, data reduction, and catalog products are detailed elsewhere (e.g. \citealp{giovanelli05a}; \citealp{saintonge07a}; \citealp{haynes11a}). Columns 1-8 of Table \ref{maintable} give the \hi\ data from the ALFALFA 70\% catalog for sources in the HUDS-R and HUDS-B samples. In brief, ID numbers (column 1) are taken from the Arecibo General Catalog (AGC), and the J2000 coordinates (columns 2 and 3) are those of the identified very low surface brightness optical source. Recessional velocities (column 4) are measured at the center of the \hi\ line at the 50\% flux level. W$_{50}$ (column 5) is the width of the \hi\ line measured at the 50\% flux level, corrected for channel broadening. \hi\ line fluxes (column 6) are calculated from fits to the spatially integrated line spectrum, Distances (column 7) are calculated using the ALFALFA flow model, which is simply Hubble Flow at c$z>6000$~\kms; for sources in this velocity range ($\sim$2000-8000~\kms) distance uncertainties due to proper motions are $\lesssim$15\%. \hi\ masses (column 8) are calculated from the given integrated fluxes and distances assuming that the gas is optically thin. 

Of particular relevance to this paper, optical identification is done by eye, matching sources in SDSS or DSS2 images with the ALFALFA position. We emphasize this visual identification, because extended nearby sources, including low surface brightness sources, are often shredded into multiple sources, and classified as more distant objects in automated catalogs. Further, we are able to identify a likely counterpart even in cases where the catalog does not include an entry due to a failure in the fit or proximity to a star. Without a corresponding optical redshift, identification necessarily relies on a small spatial offset between the optical source and the ALFALFA position. Though rare sources have been identified at large offsets from the ALFALFA centroid \citep{cannon15a}, the average ALFALFA \hi\ centroid accuracy is $\lesssim$20\arcsec, and confirmation observations have found the identifications to be quite reliable in almost all cases with an identified optical counterpart. 

\subsubsection{Synthesis Data}
\label{data.hi.synthesis}
We observed AGC122966 and AGC 334315 with 2$\times$12h pointings with WSRT as part of exploratory observations of (almost) dark sources in the ALFALFA survey (program R13B/001; PI Adams). 
The observations were centered on the central \hi\ velocity measured in ALFALFA, with a 10~MHz bandpass with 1024 channels and 2 polarizations, leaving ample line-free channels for continuum subtraction, but still sufficient velocity resolution of 4.1~\kms\ after Hanning smoothing. 
  
We observed AGC 219533 under a separate program (14B-243; P.I. Leisman) with the Karl G. Jansky Very Large Array (VLA) in 2014.
We observed the source for two 3 hour observing blocks in C-configuration, using the WIDAR correlator in dual polarization mode with a single sub-band 8~MHz wide with 1024 channels, giving a native channel width of 1.7~\kms. 

We reduced the WSRT data following the same process described in \citet{janowiecki15a} and \citet{leisman16a}, using the automated pipeline of MIRIAD \citep{sault95a} data software wrapped with a Python script (see \citealp{serra12a}; \citealp{wang13a}). The pipeline automatically 
flags the data for RFI and iteratively deconvolves the data with the
CLEAN algorithm after primary bandpass calibration, in order to apply a self-calibration. The calibration solution and continuum subtraction are applied in the visibility domain before inverting the data. The resulting data cubes are iteratively cleaned down to their rms noise, using CLEAN masks determined by clipping after filtering with Gaussian kernels.
This process produces cubes with three different robustness weightings, r=0.0, r=0.4, 
and r=6.0, and bins the data to a velocity resolution of 6.0~\kms\ after Hanning smoothing. 

We reduced the VLA data using standard procedures in the CASA package (Common Astronomy Software Applications; \citealp{mcmullin07a}), including flagging of the visibilities, calibration, and continuum subtraction. We imaged the calibrated uv data following standard procedures, producing data cubes using the CLEAN task in CASA, with a Briggs robust weighting of 0.5. Since we expected the source to be extended, we used the multiscale clean option which  improves localization of extended flux \citep{cornwell08a}. It models the source as a collection of point sources and Gaussians of the beam width and several times the beam width. 

For each source we created \hi\ total intensity maps by creating a 3$\sigma$ mask on images smoothed to 2x the beam
size, applying the mask to the unsmoothed cubes, and then summing along the velocity axis. We convert these maps to \hi\ column densities assuming optically thin \hi\ gas that fills the beam, 
and also produce \hi\ moment one maps (representing velocity fields) from the masked cubes.
The resulting \hi\ images and velocity maps are shown in Figure \ref{overlays}. 

\subsection{Optical Data}
\label{data.opt}

\subsubsection{Archival SDSS Data}
\label{data.opt.sdss}

We obtained calibrated, background subtracted SDSS optical images in the g, r, and i bands for the full sample from the SDSS mosaic server described in \cite{blanton11a}. They estimate that the uncertainty in the background contributes a systematic uncertainty of up to $\sim$10\%.  

Since the inclination is poorly constrained for the low surface brightness sources in question, we measure the average flux in concentric circular annuli to approximate the surface brightness profile of the sources, using Python code we developed based on Astropy \citep{astropy13a}, and its affiliated package Photutils. We note that we chose the galaxy center to be the center of the extended optical flux, which, for sources with clumpy morphologies and significant evidence of star formation, may not be the location of the peak flux. We then fitted exponential functions to the surface brightness profiles, including a term to estimate a constant offset due to the background. 

We correct our measured profiles for galactic extinction, but do not correct for cosmological surface brightness dimming for consistency with other local universe studies, and since the dimming corrections are small at the distances of our sample. We model the effect of the PSF on our fitted values by simulating model 1D profiles convolved with a 1D approximation of the SDSS PSF, and then applied our fitting method to both the true and convolved profiles to calculate analytic approximations of its effect. We then correct our measured parameters accordingly. We note that the sources in this sample are very extended relative to the SDSS PSF, so these corrections tend to be small.

We report the results of these fits in columns 9-12 of Table \ref{maintable}. Specifically, column 9 gives the measured central surface brightness from the exponential fit to the g-band data, and errors that are the quadrature sum of uncertainties from the fit and an assumed 10\% uncertainty in the absolute background calibration. They do not account for additional systematic errors introduced from uncertainty in the inclination or galaxy centroid. Column 10 gives the effective radius, which is 1.68$\times$ the disk scale length  from the exponential fit, and contains half the light from the galaxy. Column 11 gives the estimated absolute magnitude derived from integration of the exponential fit and the assumed distance. We note that this total magnitude has not been truncated and should be used with caution: it is significantly brighter than measurements derived from aperture magnitudes (the average offset is $\sim$0.5 mag), especially given the large estimated disk scale length for these extended sources. Column 12 gives the g-r color derived from the offset between the exponential fits, which is a better measurement than from using the absolute magnitude of these sources.   

\subsubsection{WIYN pODI Data}
\label{data.opt.podi}

We observed AGC~122966 and AGC~334315 in October 2013 with the WIYN\footnotemark[1]
3.5-m telescope at Kitt Peak National Observatory\footnotemark[2] using the partially populated One
Degree Imager (pODI). At that time, pODI had a field of view of 24\arcmin $\times$
24\arcmin, and we used a dithering sequence to eliminate chip gaps. We
obtained nine 300s exposures in SDSS $g$ and $r$ filters (\citealp{gunn98a},
\citealp{doi10a}) for both 
targets. Due to unfavorable weather conditions, we did
not observe AGC~219533. While faint, it is significantly
detected in SDSS images and we use those in this analysis.

\footnotetext[1]{The WIYN Observatory is a joint facility of the University
of Wisconsin-Madison, Indiana University, the University of Missouri,
and the National Optical Astronomy Observatory.}

\footnotetext[2]{Kitt Peak National Observatory, National Optical Astronomy
Observatory, which is operated by the Association of Universities
for Research in Astronomy (AURA) under cooperative agreement
with the National Science Foundation.}

We reduced our observations using the QuickReduce
(QR, \citealp{kotulla14a}) data reduction pipeline via the ODI Pipeline,
Portal, and Archive (ODI-PPA; \citealp{gopu14a}, \citealp{young14a}) at
Indiana University. The QR pipeline removes instrumental signatures
including bias, dark, flat, pupil ghost, nonlinearity, cosmic rays,
and fringes. We also applied an iterative dark-sky illumination correction to
produce very flat final stacked images, following the methods of
\cite{janesh15a} and \cite{janowiecki15a}, using the ``odi-tools" package\footnotemark[3].
 Our final images are calibrated
using catalog fluxes of SDSS stars in the frames \citep{alam15a} and
our photometric zeropoints typically have rms errors $\le$0.05~mag.

\footnotetext[3]{\url{https://github.com/bjanesh/odi-tools}}

\subsection{Archival UV Data}
\label{data.uv}

A subset of the HUDS samples fall within the footprint 
of archival GALEX near ultraviolet (NUV) and far ultraviolet (FUV) observations (\citealp{martin05a}; \citealp{morrissey07a}). The FUV bandpass ranges from 1344 to 1786~\AA\ with a PSF of 4.3\arcsec\ FWHM, and the NUV covers 1771 to 2831~\AA\ with a 5.3\arcsec\ FWHM. These bands are sensitive to the hard radiation from young stellar populations, and are thus useful in understanding recent star formation in these sources. 

Specifically, \cite{huang12b} studied a sample of the 9417 galaxies in the 40\% ALFALFA catalog with overlapping SDSS and GALEX coverage, and found SFRs and stellar masses via SED fitting for the sources in their ALFALFA-SDSS-GALEX sample. This includes 30 of the HUDS-B galaxies (the HUDS-BG sample). To avoid introducing systematic trends due to differences in methodology, we restrict our comparisons of stellar masses and SFRs to the larger ALFALFA-SDSS-GALEX sample to these 30 sources, but note that this is a sufficient quantity to understand trends in the HUDS-B sample. Importantly, the sources with available GALEX data - the HUDS-BG sample - 
have the same distribution of observed properties (color, \hi\ mass, absolute magnitude) as the HUDS-B sample. 
Thus, any analysis using only the HUDS-BG sources does not introduce a selection bias into our results.

Additionally, the three sources with \hi\ synthesis observations are not in the 40\% ALFALFA catalog analyzed by \cite{huang12b}, but do fall within the GALEX footprint. AGC~219533 is clearly detected in medium-deep imaging. 
AGC~122966 and 334315 are only covered in the much shallower AIS survey, and thus are only marginally detected in smoothed images. 
We use these data to roughly estimate star formation rates for these sources using standard prescriptions from \cite{kennicutt12a}, and report the results in Table \ref{synthesistable}.

\subsection{A Note on Inclinations}
\label{data.inc}

Analysis of surface brightness, surface density, and rotational parameters depends in part on the source inclination. However, optical measurements of inclinations in star forming galaxies are difficult without clear near infrared detections of the older stellar populations, an issue compounded by the low S/N of HUDS in SDSS data. 

Thus, we approach this issue in two ways. For the three sources with resolved \hi\ data we estimate inclinations from the \hi\ images, using the standard formula:
$$cos^2(i) = \frac{(b/a)^2-q_0^2}{1-q_0^2}$$
assuming that the gas forms a circular disk with an intrinsic axial ratio of $q_0$=0.2, and report the values in Table \ref{synthesistable}. For our sources, the dominant error in this calculation comes from the size of the \hi\ beam; we calculate uncertainties assuming errors of half the beam width along the kinematic major and minor axes. This uncertainty contributes significantly to our estimates of dynamical masses and spin parameters, but is still well enough constrained to provide
 significant constraints.

For sources without \hi\ synthesis data, we assume that all sources are inclined at 45$^{\circ}$ for purposes of measuring rotational velocities and spin parameters, and we do not correct our surface brightness measurements for inclination. We assess the impact of these assumptions on our calculated distributions and sample selection by repeating the calculations assuming inclinations measured from the SDSS catalog (which are very uncertain for the HUDS). 
We find no significant differences in our results, and that our measured central surface brightnesses are consistent with those measured by \cite{vandokkum15a} within our errors of $\sim$0.2~mag~arcsec$^{-2}$.

\section{Results}
\label{results}

\begin{table*}[]
\caption{\textbf{Derived Properties of HUDs with Resolved \hi\ Imaging}}
\vskip 5pt
\centering
\footnotesize{
\begin{tabular}{lcccccccccc}
\hline
\hline
AGC ID & log(M$_{\star})^a$ & SFR$^b$   & M$_{\rm HI}$/M$_{\star}$ &  D$_{\rm max}^c$ & D$_{HI}^d$ & D$_{\rm HI,pred}^e$ & S$_{\rm syn}$/S$_{\rm ALFA}^f$  & $i^g$ & M$_{\rm Dyn, 8kpc}^h$ & M$_{\rm Dyn, Max}^i$\\
       &   log M$_{\odot}$ & \msun/yr &   & kpc  & kpc  & kpc &  & deg &  10$^9$ M$_{\odot}$ & 10$^{10}$ M$_{\odot}$ \\
\hline
  122966 & 8.1  & 0.022   & 8.3  & 44$\pm$7 & 21$\pm$7  & 18$\pm$2  & 1.08  & 52$\pm$19  & 5.1$\pm$3.2 & 1.0$\pm$0.6 \\
  219533 & 7.8  & 0.030   & 24  & 38$\pm$3  &  27$\pm$3 & 24$\pm$2  & 0.96  & 47$\pm$11   & 10.0$\pm$4.1 & 1.9$\pm$0.7 \\
  334315 & 7.8  & 0.045   & 23  & 38$\pm$6  &  26$\pm$6 & 22$\pm$2  &  0.97  & 52$\pm$9   & 5.2$\pm$1.4 & 1.0$\pm$0.3 \\
\hline
\hline
\end{tabular}\label{synthesistable}
\begin{flushleft}
$^a$: Stellar mass calculated from in-house photometry on SDSS images and using the relations from \cite{zibetti09a}; typical errors are 0.3-0.4 dex. $^b$: Star formation rate estimated from the FUV GALEX luminosity using the standard relation of \cite{kennicutt12a}; errors are $\sim$50\%. $^c$: Maximum extent of \hi\ emission on the sky, uncorrected for beam smearing. $^d$: \hi\ diameter measured from 1D RADIAL models at a surface density of 1\msun$~pc^{-2}$.  $^e:$ Predicted \hi\ diameter from \cite{broeils97a}. $^f:$ Ratio of flux recovered in WSRT or the VLA data versus the measured ALFALFA flux. $^g:$ The inclination, as measured from \hi\ synthesis data. $^h:$ Dynamical mass within 8~kpc estimated from the \hi\ velocity field. $^i:$ Dynamical mass estimated from maximum observed \hi\ rotational velocity. 
\end{flushleft}
}
\end{table*}

\begin{figure*}
\centering
{\bf \large \hspace{0.3in} AGC219533 \hspace{1.05in} AGC~122966 \hspace{0.98in} AGC~334315}\\
\includegraphics[width=0.31\textwidth]{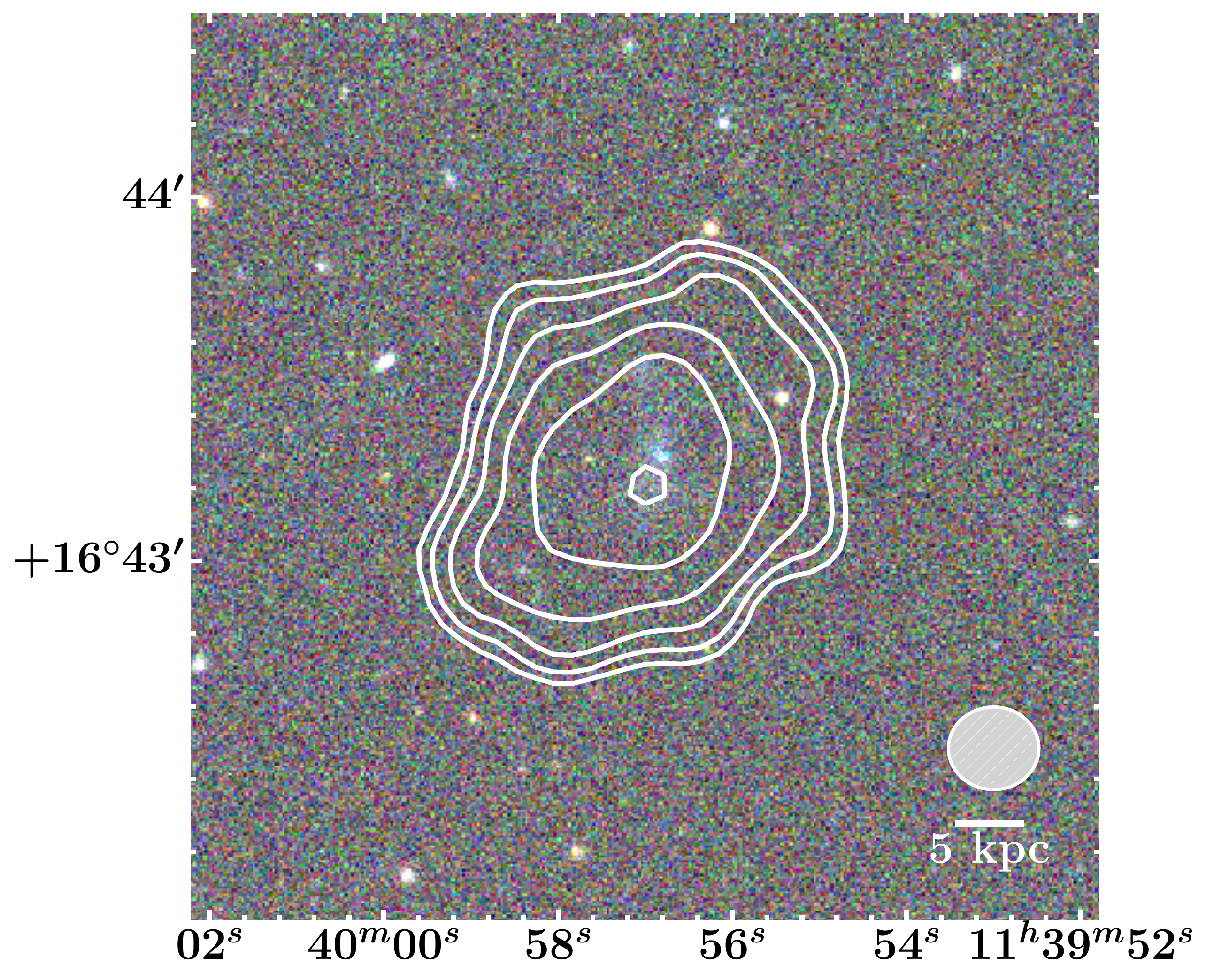}
\includegraphics[width=0.3\textwidth]{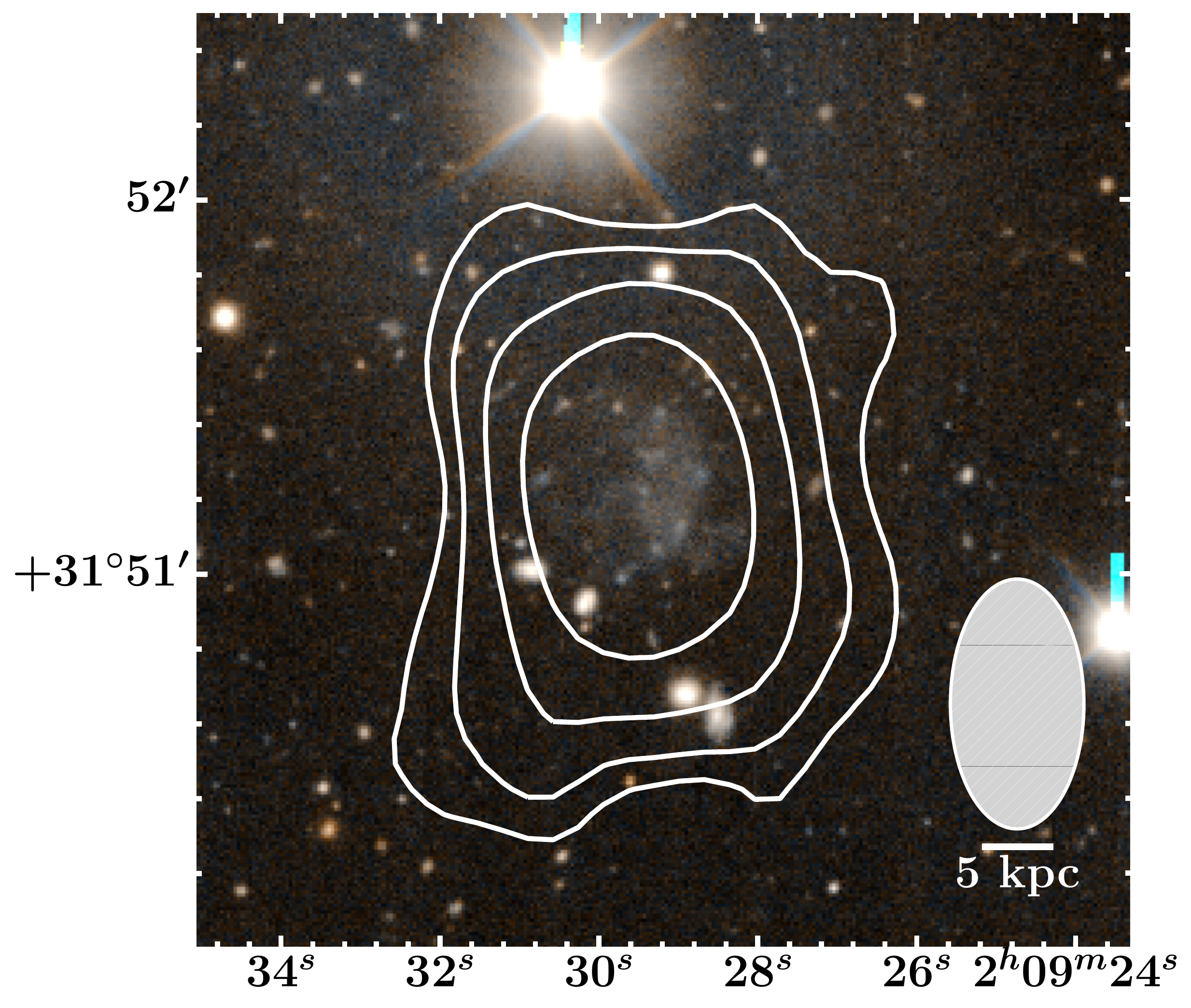}
\includegraphics[width=0.293\textwidth]{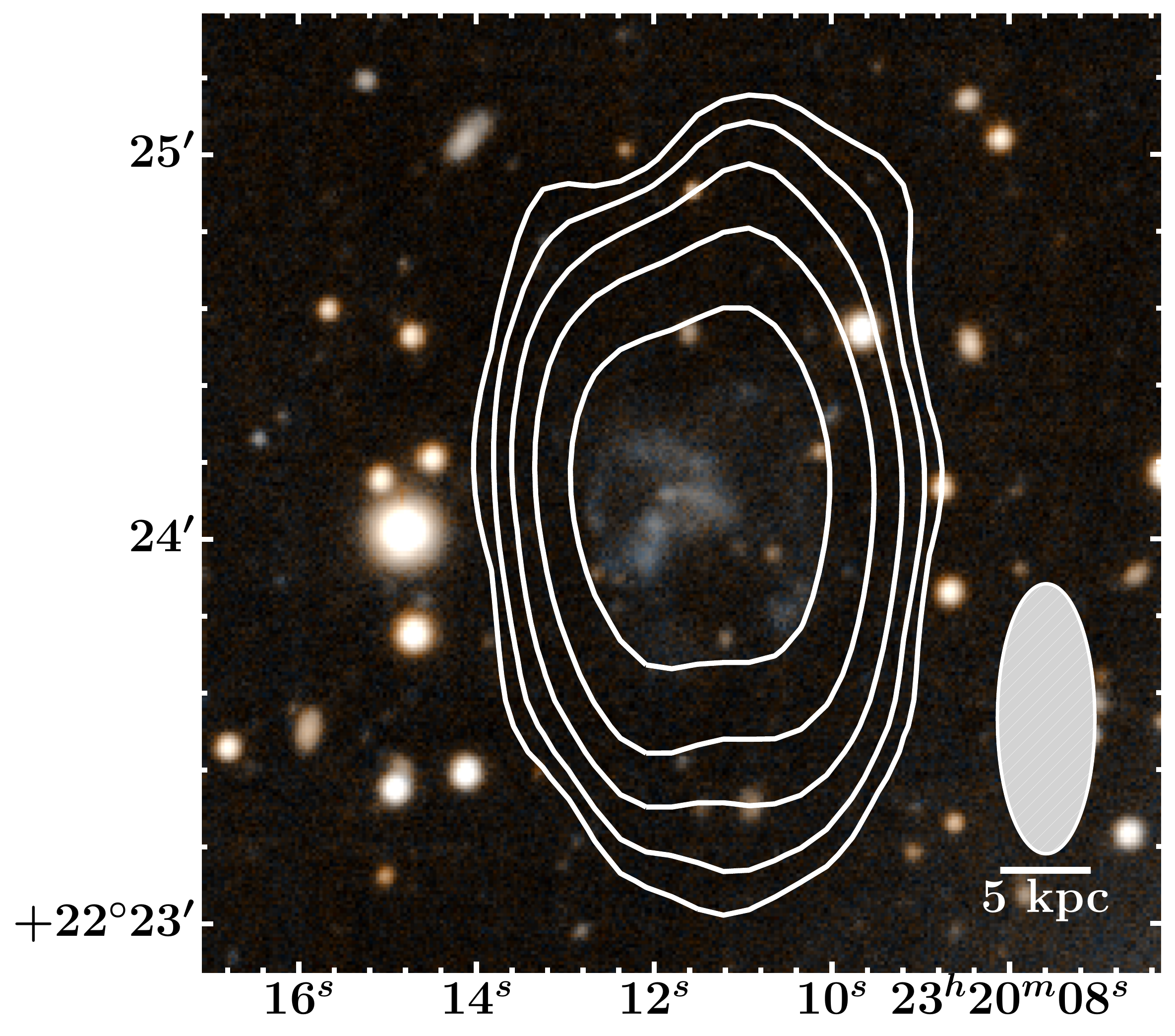}\\
\includegraphics[width=0.3\textwidth]{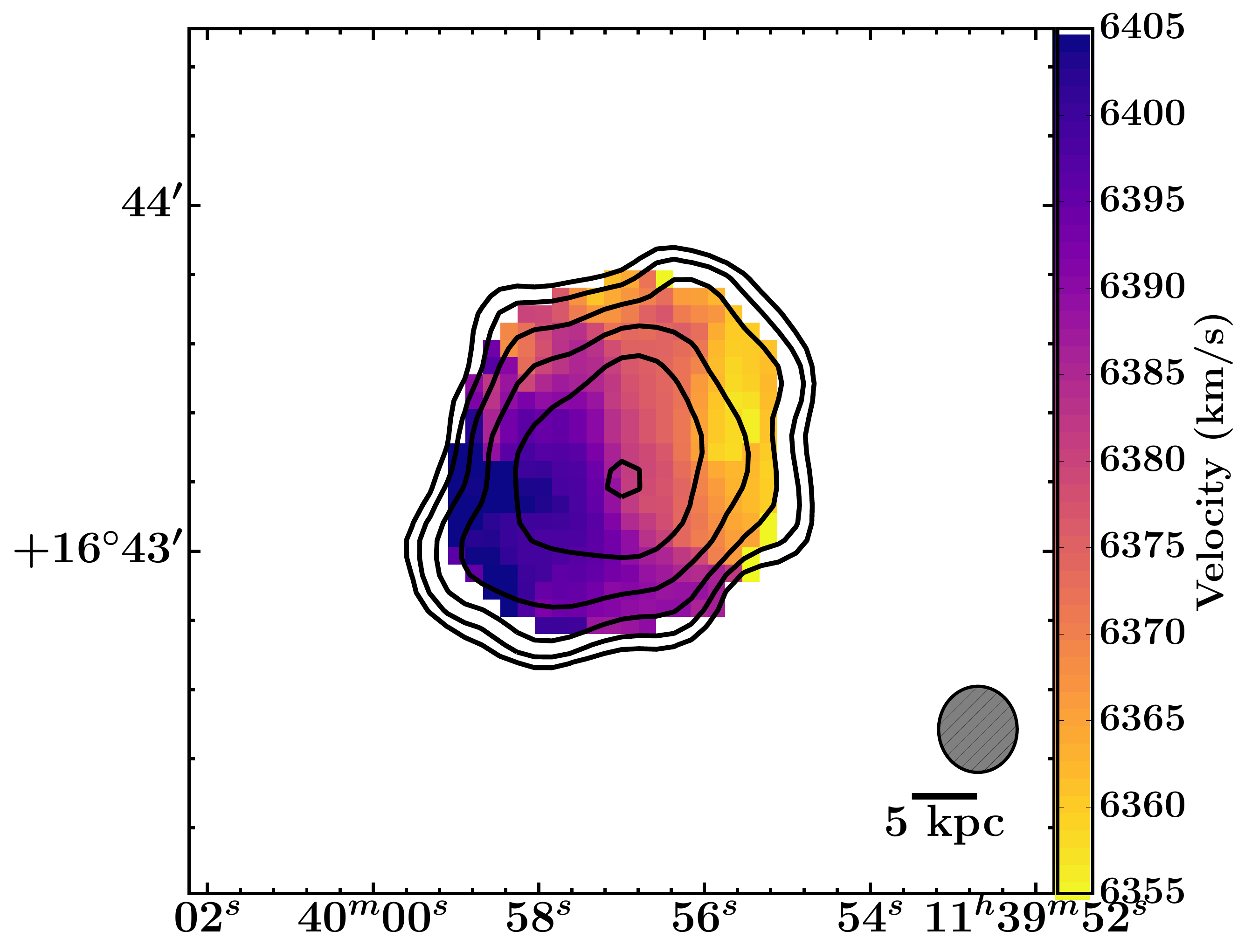}
\includegraphics[width=0.3\textwidth]{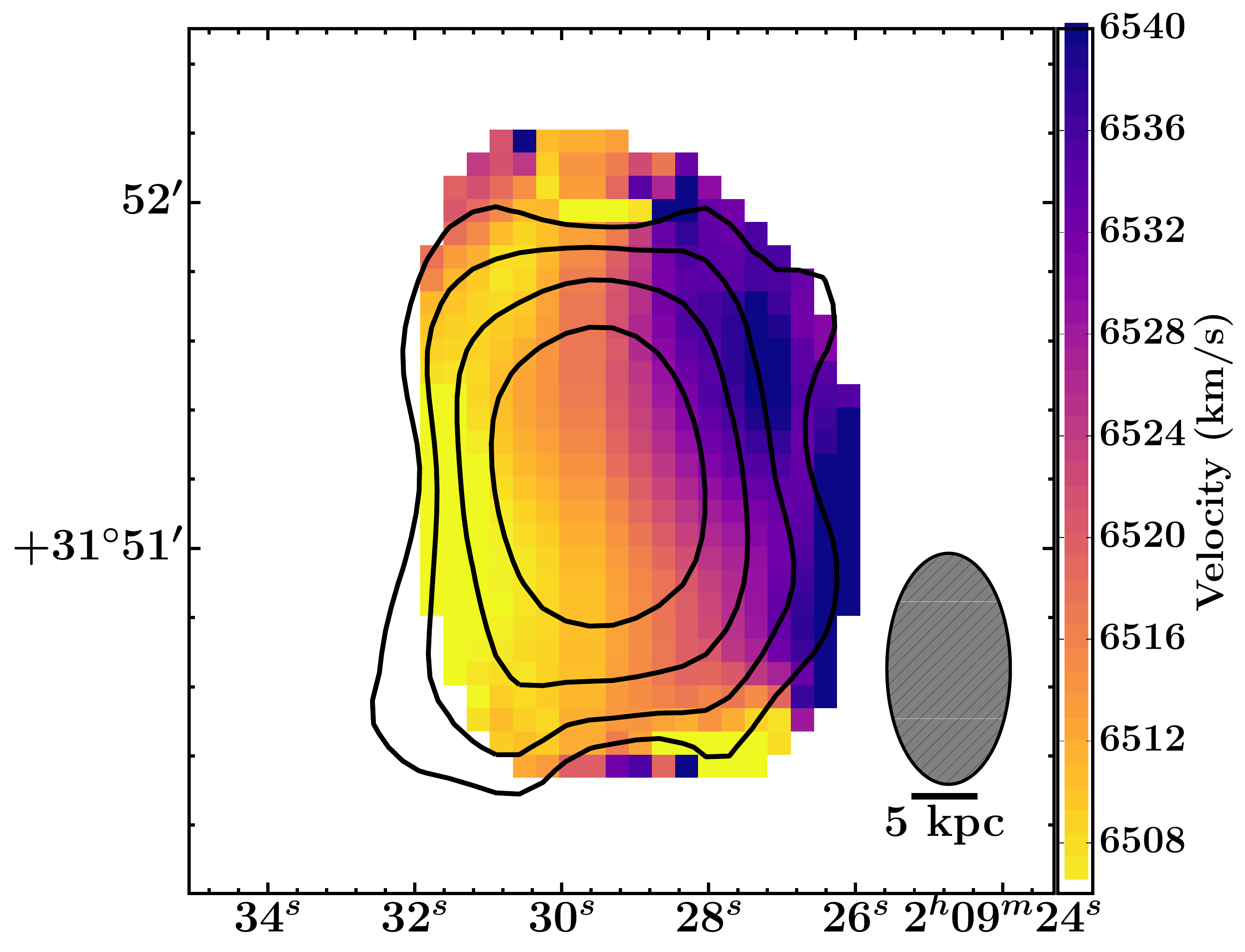}
\includegraphics[width=0.3\textwidth]{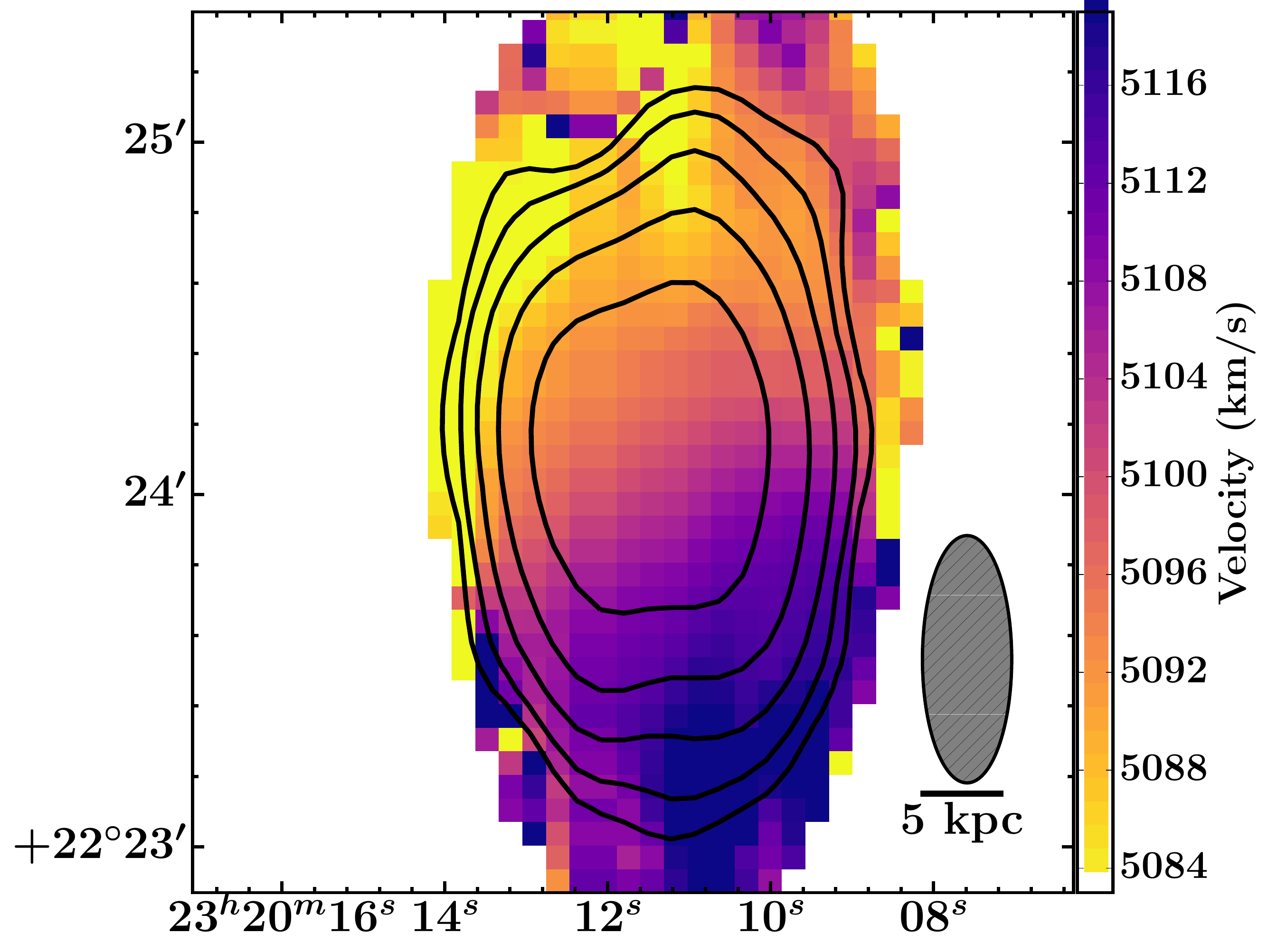}
\vspace{-0.3cm}
\caption{\hi\ column density contours at 2, 4, 8, 16, 32, and 64$\times$10$^{19}$~atoms~cm$^{-2}$ overlaid on color optical images and moment 1 velocity maps for three HUDS, AGC~219533, 122966, and 334315 from left to right. For AGC~219533 (far left) the \hi\ data is VLA C-array and the optical data is from SDSS; the others have \hi\ data from WSRT, and optical data from pODI on the WIYN 3.5m. The optical emission is blue, diffuse, and shows irregular morphology. The \hi\ is resolved even at this low physical resolution, is significantly extended relative to the diffuse optical emission, and shows evidence of ordered rotation.  RA and Dec are in J2000 coordinates. 
\label{overlays}
}
\end{figure*}

Here we present the optical and \hi\ properties of the HUDS from the ALFALFA survey. Section \ref{results.optical} describes the optical properties of the galaxies, emphasizing that while they have similarly large extents for their stellar mass, they differ from previously reported UDGs in color and morphology. Section \ref{results.hi} describes the \hi\ properties of the sources, emphasizing their large \hi\ masses and diameters given their stellar mass.   

\subsection{Optical Properties}
\label{results.optical}
\begin{figure}
\centering
\includegraphics[width=\columnwidth]{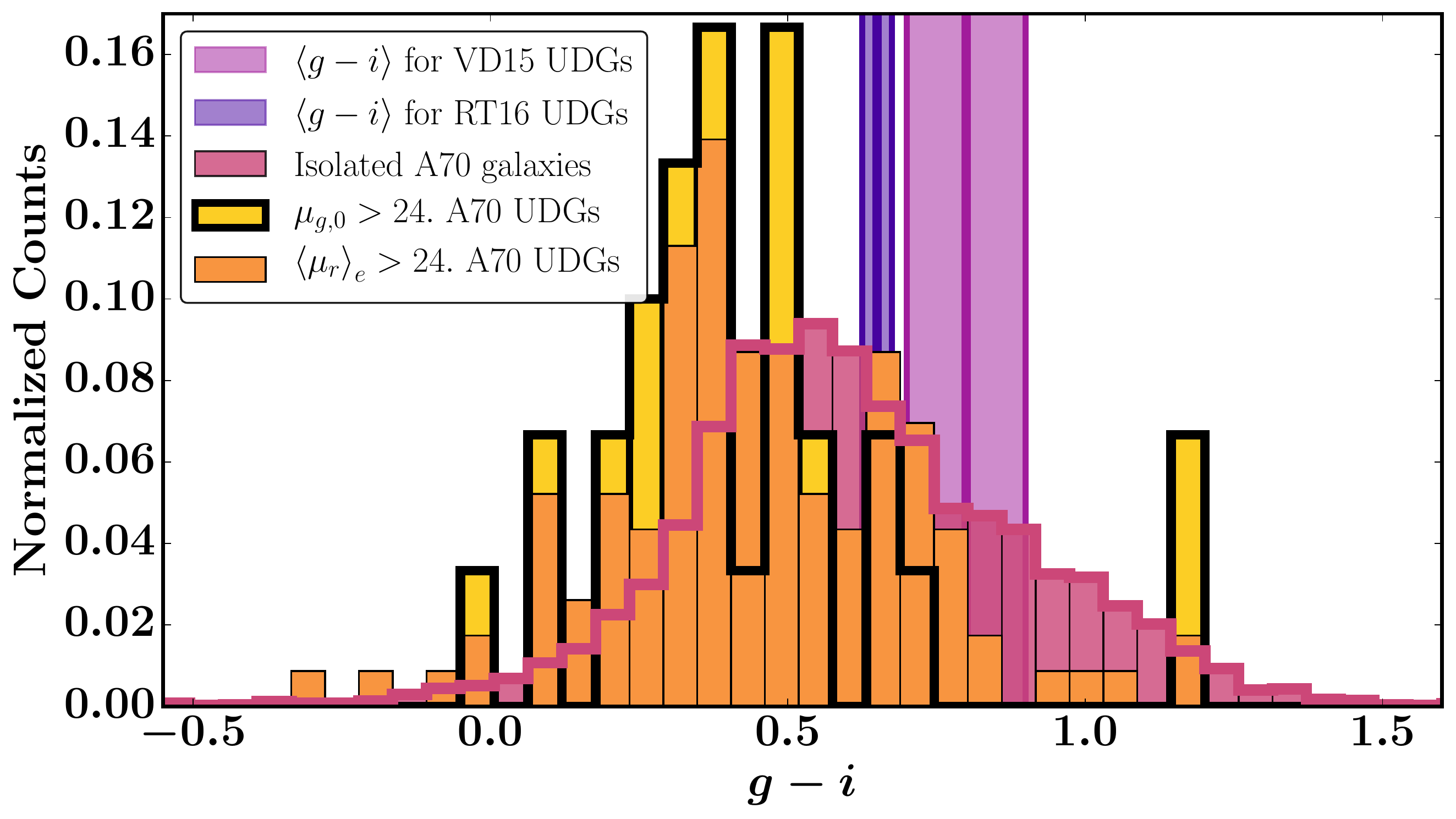}\\
\includegraphics[width=\columnwidth]{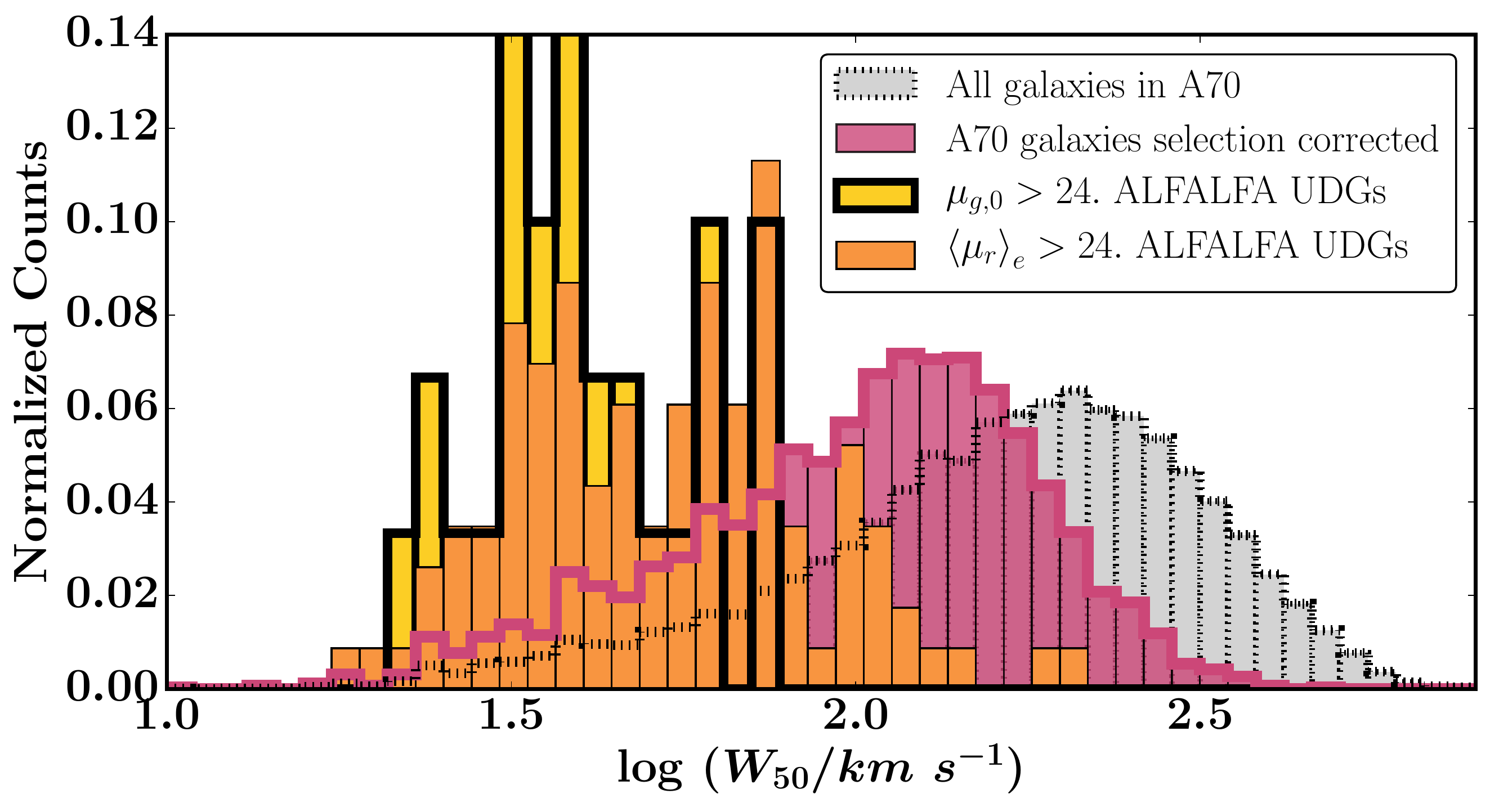}
\includegraphics[width=\columnwidth]{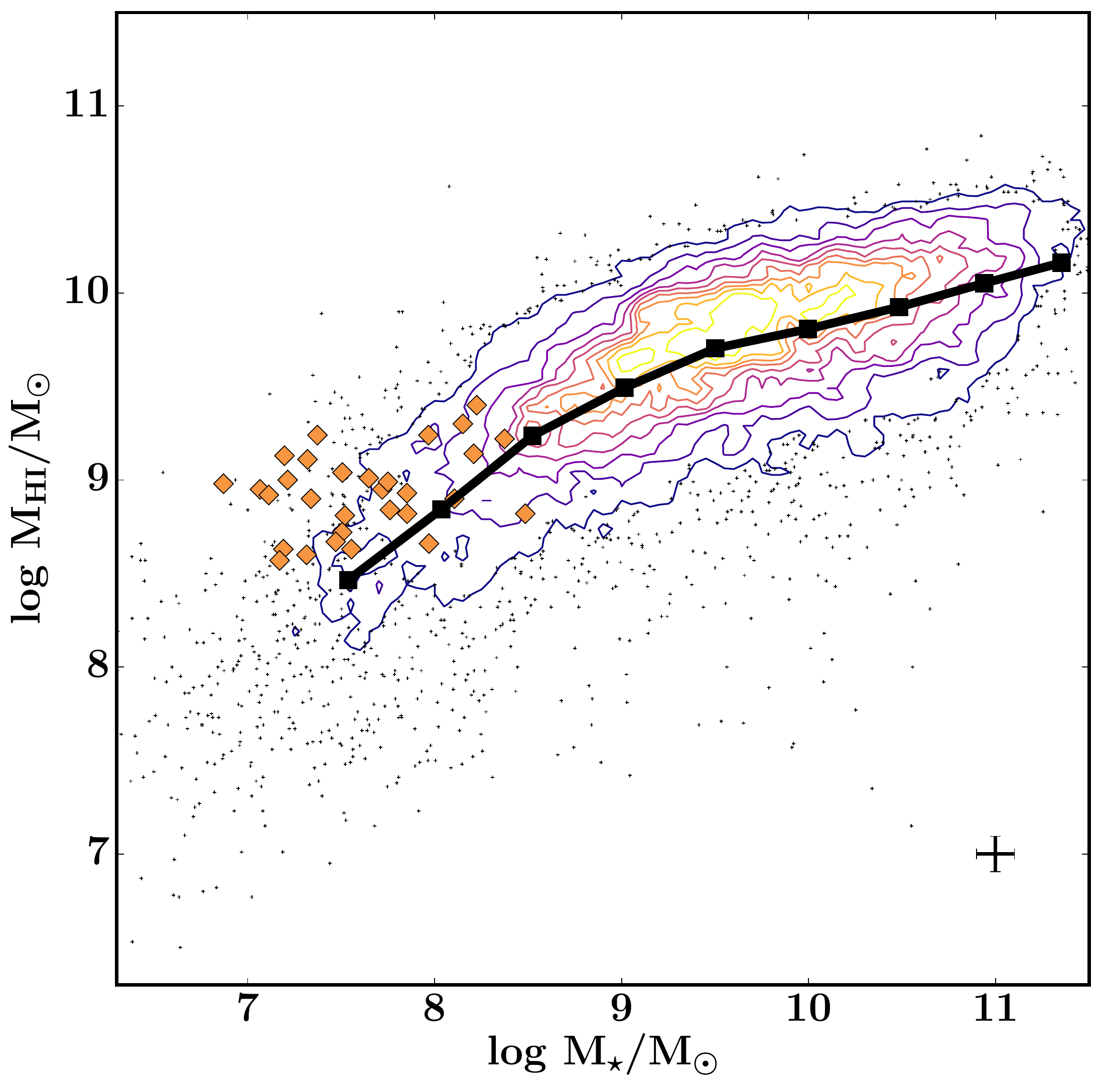}
\vspace{-0.1cm}
\caption{Top: The color distribution of HUDS (yellow and orange for HUDS-R and HUDS-B respectively), compared with the rest of the ALFALFA population matching similar isolation and distance selection criteria, the average value computed from stacking in \cite{vandokkum15a}, and the average value for the sources in filaments from \cite{roman16a}. Middle: The distribution of \hi\ velocity widths as measured at the 50\% flux level of HUDS (yellow and orange for HUDS-R and HUDS-B respectively), compared with the rest of the ALFALFA population (light grey), and the the ALFALFA population corrected for inclination and mass selection effects (pink). HUDS tend to have quite narrow velocity widths, even when correcting for selection effects. Bottom: \hi\ mass versus stellar mass for ALFALFA sources from \cite{huang12b}, compared with the HUDS-BG sample. HUDS tend to be \hi-rich relative to their stellar mass.
\label{distributions}
}
\end{figure}

The top panels of Figure \ref{overlays} show SDSS and ODI color images of three HUDS with synthesis follow up observations. Comparing these images with those of, e.g., the Coma UDG DF~17 pictured in Figure \ref{SDSSodiCompare}, emphasizes the morphological difference between the HUDS and other UDGs. While DF~17 shows a smooth stellar distribution even in deep CFHT imaging, deep WIYN imaging of two HUDS show clumpy, irregular morphologies, with knots of comparatively intense star formation. AGC~122966 shows two arcs superimposed on extended faint emission, while AGC~334315 similarly shows brighter arcs crisscrossing fainter extended emission. Both stellar populations appear quite disturbed, with the peak surface brightness offset from the center of the low surface brightness emission. 

These morphological differences can have implications for the definition of ``ultra-diffuse." That these sources are significantly extended and very low surface brightness is clear: SDSS and pODI imaging measure diameters ranging from 24 to 70 arcseconds, which translates to diameters between 11 and 25~kpc at their respective distances. However, profile fitting in the traditional sense is complicated by the lack of a smooth profile. While the central surface brightness measured from profile fitting corresponds well with the source peak surface brightness for smooth, quiescent sources, the peak surface brightness may be offset from the center of light in these patchy sources, making estimates of the surface brightness profile somewhat sensitive to the chosen aperture center. 

The color images in Figure \ref{overlays} also emphasize the blue nature of the stellar population of HUDS. The top panel of Figure \ref{distributions} shows the color distribution of the HUDS-B and HUDS-R samples compared with other ALFALFA galaxies that meet matching distance and isolation criteria, and the average color estimate from \cite{vandokkum15a}. The average g-i color of HUDS-B is 0.45, with a standard error of 0.02, significantly bluer than the 0.8$\pm$0.1 estimated by \cite{vandokkum15a}, and also bluer than the 0.65$\pm$0.02 estimated by \cite{roman16a} for UDGs outside of clusters. 
The color appears only slightly bluer than the color distribution of star forming ALFALFA sources within a similar mass range (see Figure \ref{distributions}), suggesting that their color is more directly tied to their \hi\ than to their diffuse stellar distribution. 

These differences in color again have important implications for the definition of an ``ultra-diffuse" population. For example, in order to appropriately make the comparison between the samples plotted in Figure \ref{compare} we have converted sources observed in g and r to V band, which falls between the g and r filters. Thus, in plots made with g band values the HUDS would shift to brighter values relative to the UDG population, and would shift to fainter values for plots in r band.

A second, more striking implication, however, is that if the recently formed (bright, blue) stars were not present in these sources, the remaining stellar populations would be significantly fainter. Thus, the older stellar populations of optically selected UDGs are likely significantly brighter than any (currently invisible) older stellar populations in these HUDS. Thus, though they are similar sources in terms of measured parameters, these sources may have significant physical differences from other UDGs. However, their low surface brightness nature still implies a connection: it may be that these sources are progenitor UDGs, fainter and less evolved versions of their more evolved cluster counterparts.

\subsection{HI Properties}
\label{results.hi}

In contrast with optically selected UDGs in denser environments, the isolated HUDS are clearly detected in \hi, with \hi\ masses ranging from 10$^{8.6}$-10$^{9.3}$\msun. In fact, these sources are gas rich, even relative to the normal gas-bearing galaxy population. The bottom panel of Figure \ref{distributions} shows the \hi\ mass - stellar mass scaling relation for \hi\ selected galaxies from \cite{huang12b}. For a given stellar mass, the HUDS have fairly high gas fractions relative to the rest of the ALFALFA population, similar to fainter and smaller dwarf irregulars (e.g., \citealp{lee03a}), and pushing into a similar parameter space to (almost) dark sources like those reported in \cite{janowiecki15a}. This makes sense given our selection criteria: the minimum distance threshold eliminates sources with \hi\ masses $\lesssim$10$^8$\msun, and our absolute magnitude limit places a stellar mass threshold of $\sim$10$^9$\msun. However, while this selection eliminates sources with gas fractions $<$0.1, the mean gas fraction M$_{\rm HI}$/M$_*$=35 may suggest a potential connection between high gas fraction and the diffuseness of the stellar population. Regardless, the \hi\ dominates the baryonic content of these galaxies. Whether we interpret them in terms of their stellar mass or their baryonic mass thus makes a significant difference, a point we return to in section \ref{discussion}.

\subsubsection{The \hi\ Distribution}
\label{results.hi.dist}

While these HUDS appear to have large \hi\ masses relative to their stellar mass, UDGs are optically defined by their large radii for their stellar masses. Thus,
for the three sources with existing \hi\ synthesis observations we analyze their \hi\ radii and distribution. Estimates of their properties are limited by the low physical resolution of the data (6 - 14~kpc), but are still sufficient to constrain their extended nature. 

All three sources are resolved with multiple beams in \hi, which allows us to estimate the radii of the sources, albeit with a fairly large uncertainty.  As a first order estimate, we place an upper limit on the size of each source by measuring the largest projected extent on the sky, uncorrected for the effects of beam smearing. Specifically we measure largest extents of 44$\pm$7, 38$\pm$3, and 38$\pm$6~kpc for AGC~122966, 219533, and 334315 respectively, assuming uncertainties of half the beam width along the major axis.
We next estimate radii by fitting the observed \hi\ profile with tilted rings every half beam width using the GIPSY task ELLINT, assuming a constant position angle and inclination, and then estimating radii at a surface density of 1~\msun~pc$^{-2}$ to compare to measurements from The \hi\ In Nearby Galaxies Survey (THINGS; \citealp{walter08a}) described below. We choose the major axis to be 
the kinematic major axis (see section \ref{results.hi.rot}), which approximately corresponds to the morphological major axis, except in the case of AGC~122966, where the morphological major axis is poorly determined due to the elongated WSRT beam.  

However, fitting flux in rings is limited by the minor axis resolution, so we additionally estimate the surface density profile using Lucy-Richardson deconvolution (\citealp{lucy74a}; \citealp{warmels88c}), which collapses the flux to a 1D profile along the major axis, and then models that profile as a disk of uniform coplanar rings. This method has the advantage of not requiring an estimate of the inclination of the sources, and is insensitive to low resolution along the minor axis, but is still limited by the resolution along the major axis. Outside of the central beamwidth, \cite{warmels88c} estimate the uncertainty in the modeling as $\sim$25\%. The surface density profiles estimated from the Lucy method are consistently higher than those from the 2D modeling with ELLINT by $\sim$25\%, but the measured radii are consistent within half the beam width, our estimated error. 

We then use these radial models to estimate the radii at 1~\msun~pc$^{-2}$ to roughly compare to the expected radii from standard scaling relations. Using the \hi\ mass-radius relation from \cite{broeils97a} (which is similar to the relation from \citealp{wang16a}) we compute expected \hi\ diameters from the measured \hi\ masses. The ratios of the measured diameters to the predicted diameters are given in Table \ref{synthesistable}. All of the sources lie above the relation, i.e., their \hi\ radius is extended for their \hi\ mass. However, this comparison is still limited by the effect of the beam, which tends to push flux to larger radii, exaggerating the size of a galaxy, while reducing the measured column density, which can underestimate the size of the galaxy for low density systems. Thus, 
given the uncertainties in the radius measurements, all three sources are consistent with the scatter in the relation. We note that while the \hi\ radii appear to be consistent with the expected radii for their \hi\ mass, as noted above, all three sources have large \hi\ masses relative to their stellar populations. Thus, these sources are significantly more extended than typical \hi-rich sources with comparable stellar mass.

\begin{figure}[]
\centering
\includegraphics[width=0.99\columnwidth]{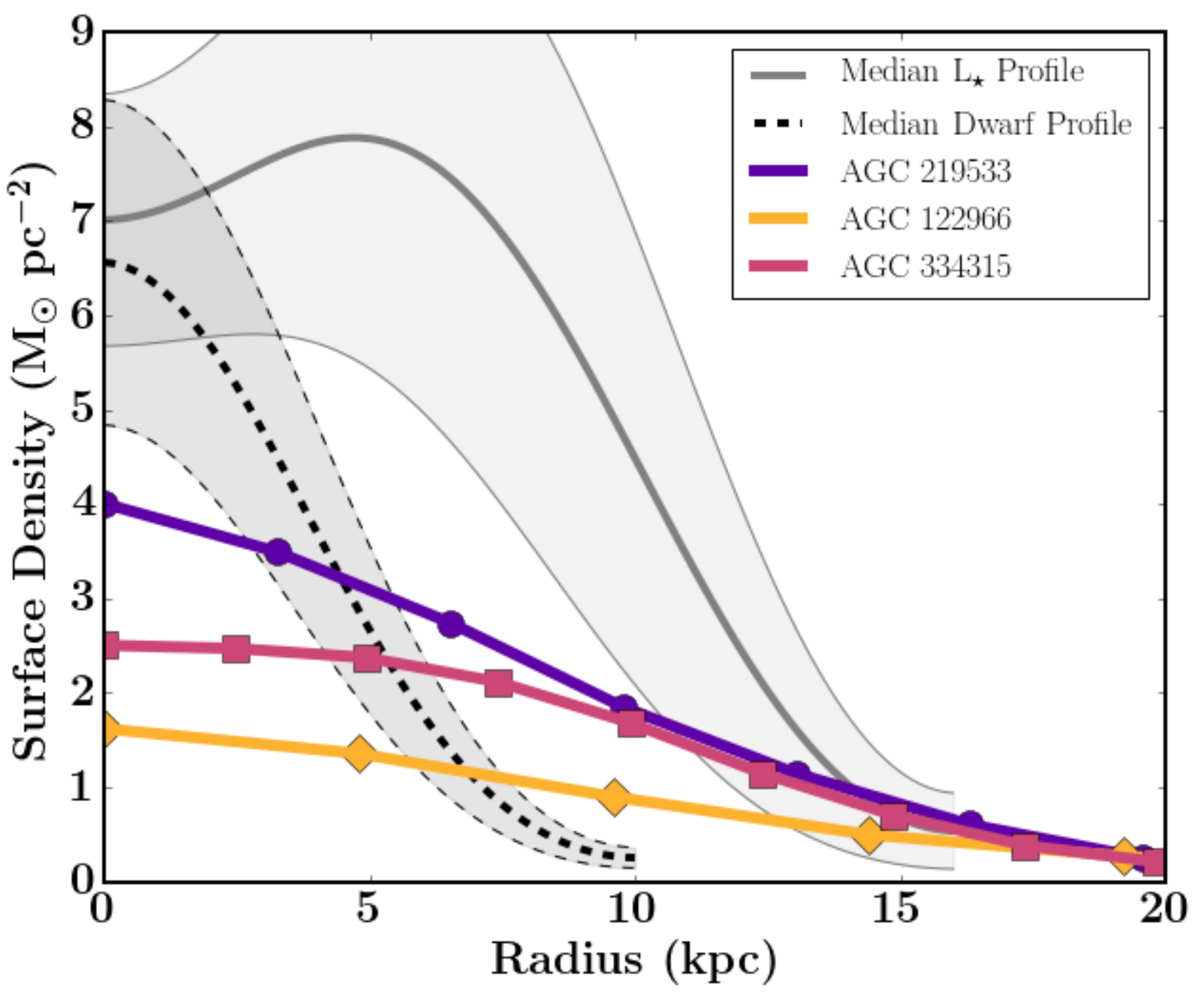}\\
\caption{Low resolution surface density profiles of HUDS with resolved \hi-synthesis imaging, compared with the median profiles of the dwarf galaxies and  $\sim$L$_{\star}$ galaxies from THINGS \citep{leroy08a}, smoothed to a physical beam resolution of 7~kpc, to match the resolution of the observed HUDS. The three HUDS are significantly more extended than the typical \hi-rich dwarf, which is approximately a point source at 7~kpc resolution, and appear to be somewhat lower column density than the typical L$_{\star}$ galaxy, though beam smearing limits interpretation of the surface density within the central beam.
\label{surfacedensity}
}
\end{figure}

As a suggestive exercise, we compute the median \hi\ profile of 4 dwarf galaxies and 4 $\sim$L$_{\star}$ galaxies from the THINGS sample \citep{walter08a}, and compare the results to the results for HUDS. We used profiles fitted using the tilted rings method from \cite{leroy08a}, and smoothed them to a physical resolution of 7~kpc to approximately compare with the physical resolution of the measured HUDS. The results of this exercise are shown in Figure \ref{surfacedensity}. The beam smeared profiles of the HUDS are shown as thick colored lines, and the smoothed median profiles and their spread are shown as grey shaded regions. Importantly, the \hi\ disks are more extended than typical dwarf galaxies which are point sources at 7~kpc resolution, and more consistent with \hi\ disks of L$_{\star}$ spirals. The average surface density seems to be somewhat lower than the typical THINGS galaxies, suggesting that these sources may be somewhat more diffuse in \hi\ than typical \hi\ sources. However, we emphasize that this result is at best suggestive due to the small number statistics and low resolution; higher resolution observations of a larger sample will be required to confirm this suggestion.

\subsubsection{The \hi\ Rotation Velocities}
\label{results.hi.rot}
The bottom panels of Figure \ref{overlays} shows the \hi\ velocity fields for the three sources with resolved synthesis observations.  All the sources show signs of ordered motions, and evidence of a significant velocity gradient, though the gradient is only over a relatively narrow range. Indeed, AGC~219533 has the largest velocity width of 66~\kms. The relatively narrow velocity widths of the three resolved sources, however, are consistent with the ALFALFA velocity widths for the entire \hi-bearing UDG sample. The center panel of Figure \ref{distributions} shows the velocity distribution of HUDS compared with a similarly selected ALFALFA sample. The mean velocity width of the HUDS-B (HUDS-R) sample is 59 (44)~\kms, compared to 194~\kms\ for all ALFALFA galaxies, and 119~\kms\ for ALFALFA galaxies with similar selection criteria. Specifically, we expect lower velocity widths for HUDS given their lower baryonic masses, and also due to the fact that surface brightness is a function of inclination. However, even after removing sources with \hi\ mass (log (\mhi/\msun)$>$9.5) and with inclinations $>$66 degrees (approximating the inclination distribution of HUDS), the HUDS still populate the low velocity width part of the distribution. This result is not entirely unexpected, due to their lower mass and a surface brightness selection against edge-on galaxies. We return to this in section \ref{discussion.dm.spin}.

Since the sources are only resolved with a few beams along the major and minor axes, traditional fitting of tilted ring models to the 2D velocity profile tend to overestimate the dispersion and underestimate the rotational velocities due to the many velocities along the lines of sight contained within the beam. Thus we instead estimate the rotation curve of the sources by fitting the ``envelope" of velocities observed at each position in a position-velocity (PV) diagram (e.g., \citealp{sancisi79a}). We follow the methods of \cite{hallenbeck14a}, using the GIPSY task ROTCUR to estimate the position angle of the velocity field (using tilted ring models), and then extract a position velocity field using a slice 2 beams wide along the major rotational axis. We then extract the spectrum at each position and fit a 3rd order Gauss-Hermite polynomial, and estimate the final rotation curve as the velocity where the integrated area under the curve is 3.3\%
 from the approaching or receding edge. 
We then average the rotation values from the approaching and receding envelope, and correct for inclination by dividing by sin(i). 

We note that the inclination uncertainty contributes significantly to the rotational uncertainty (section \ref{data.inc}), and that our estimated rotation velocity could be biased by gas inflow or outflow, enhanced velocity dispersion, and the assumptions of a disk like structure with a negligible disk scale height. In spite of these limitations, however, the data are still sufficient to constrain the allowed parameter space, as we discuss below.

\begin{figure*}[t!]
\centering
\includegraphics[width=\textwidth]{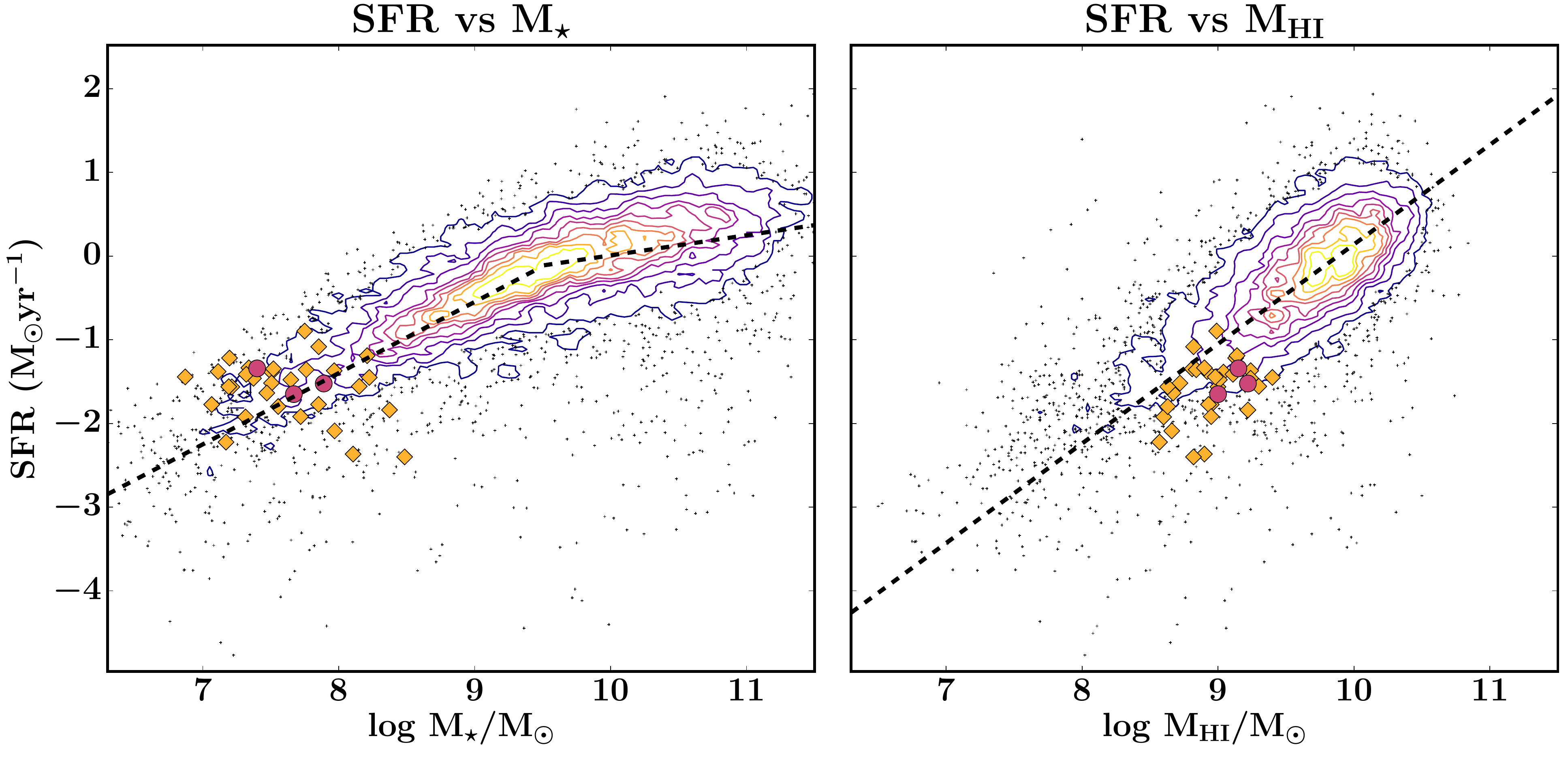}\\
\caption{SFR versus stellar and \hi\ mass of the HUDS-BG galaxies (yellow diamonds), plotted against the full \cite{huang12b} sample. The star formation rate of these sources seems typical of \hi\ selected sources of the given stellar mass, but low for galaxies with the given \hi\ mass. SFRs are taken from \cite{huang12b}, and are calculated by SED fitting to GALEX+SDSS UV broadband data. The dashed lines represent the fitted trends from \cite{huang12b}. The three HUDS with follow up observations are shown by pink circles. 
\label{sfr}
}
\end{figure*}

\section{Discussion}
\label{discussion}

In section \ref{results} we emphasized that while the HUDS have similar surface brightnesses, optical radii, and magnitudes to UDGs, they have very different colors and morphologies, are all very isolated, and all have significant \hi. Thus, their relationship to quiescent UDGs is unclear. Like other UDGs, they may be star-poor, failed $\sim$L$_{\star}$ galaxies with suppressed star formation laws, or they may be \hi-rich, extended dwarfs that only recently acquired their gas. Here we consider the star formation laws and velocity width and rotation curves of HUDS as potential evidence that they are extended dwarfs in high angular momentum halos.

\subsection{Star formation in \hi-bearing UDGs}
\label{discussion.sf}

\hi\ selected galaxies are well known to be blue, and undergoing recent star formation (e.g., \citealp{huang12b}). The HUDS are no exception. 
Figure \ref{sfr} shows the SFR versus stellar and \hi\ mass of the 33 HUDS-BG galaxies (section \ref{data.uv}) plotted against the full ALFALFA-SDSS-GALEX  sample from \cite{huang12b}. The HUDS-BG sample has moderate SFRs ranging from ~0.01-0.1 \msun~yr$^{-1}$. Indeed, in spite of their low surface brightness, the HUDS appear to have normal star formation rates for their stellar mass, i.e., their specific star formation rates are consistent with the overall ALFALFA sample. 

However, the right hand panel of Figure \ref{sfr} shows the SFRs compared with the \hi\ mass. Though the SFRs for the HUDS fall within the range covered by the ALFALFA sample, they are low compared to the average ALFALFA galaxy for a given \hi-mass, i.e., that they have very low star formation efficiency (SFE=SFR/M$_{\rm HI}$). The low SFE of these galaxies indicates that their current gas consumption time (the Roberts time, $t_R=M_{\rm HI}/SFR$) is very long, even relative to a \hi\ selected population. The average $t_R$, for the HUDS is 35~Gyr, compared to 3~Gyr for the optically selected GASS sample \citep{schiminovich10a}, and 8.9~Gyr for ALFALFA. This is not simply a selection effect: $t_R$ is nearly independent of stellar or \hi\ mass \citep{huang12b}.

There are at least two potential explanations for the long gas consumption time. 
If these galaxies continue to form stars at the same rate, they may be, in some sense, ``failed" galaxies with unusually stable disks and highly inefficient star formation. Whether they are ``failed" L$_{\star}$ galaxies or ``failed" smaller galaxies (like the Large Magellanic Cloud) depends on their estimated halo masses (see section \ref{discussion.dm}). In the latter case, these sources may be thought of as ``failed" dwarfs, and may suggest a link between the inefficiency of their star formation and their large optical radii. It may also be that these sources are selected, by means of their surface brightness and isolation, to be observed in a special time in their history. If, as \cite{dicintio16a} suggest, UDGs have bursty SF histories, we may be observing the HUDS during a period of significant gas infall, before they experience a significant increase in their star formation rate. While the resolved \hi\ imaging of these sources is smooth at the current resolution, there are not enough beams across the sources to definitively search for signs of inflow or outflow. Moreover, a moderate enhancement in SFR would make these galaxies appear brighter, and therefore too high surface brightness to be selected as ``ultra-diffuse," since observations of the older stellar populations under starbursts are difficult \citep{janowiecki14a}.

\subsection{The Dark Matter Halo}
\label{discussion.dm}

\begin{figure*}[]
\centering
\includegraphics[width=0.45\textwidth]{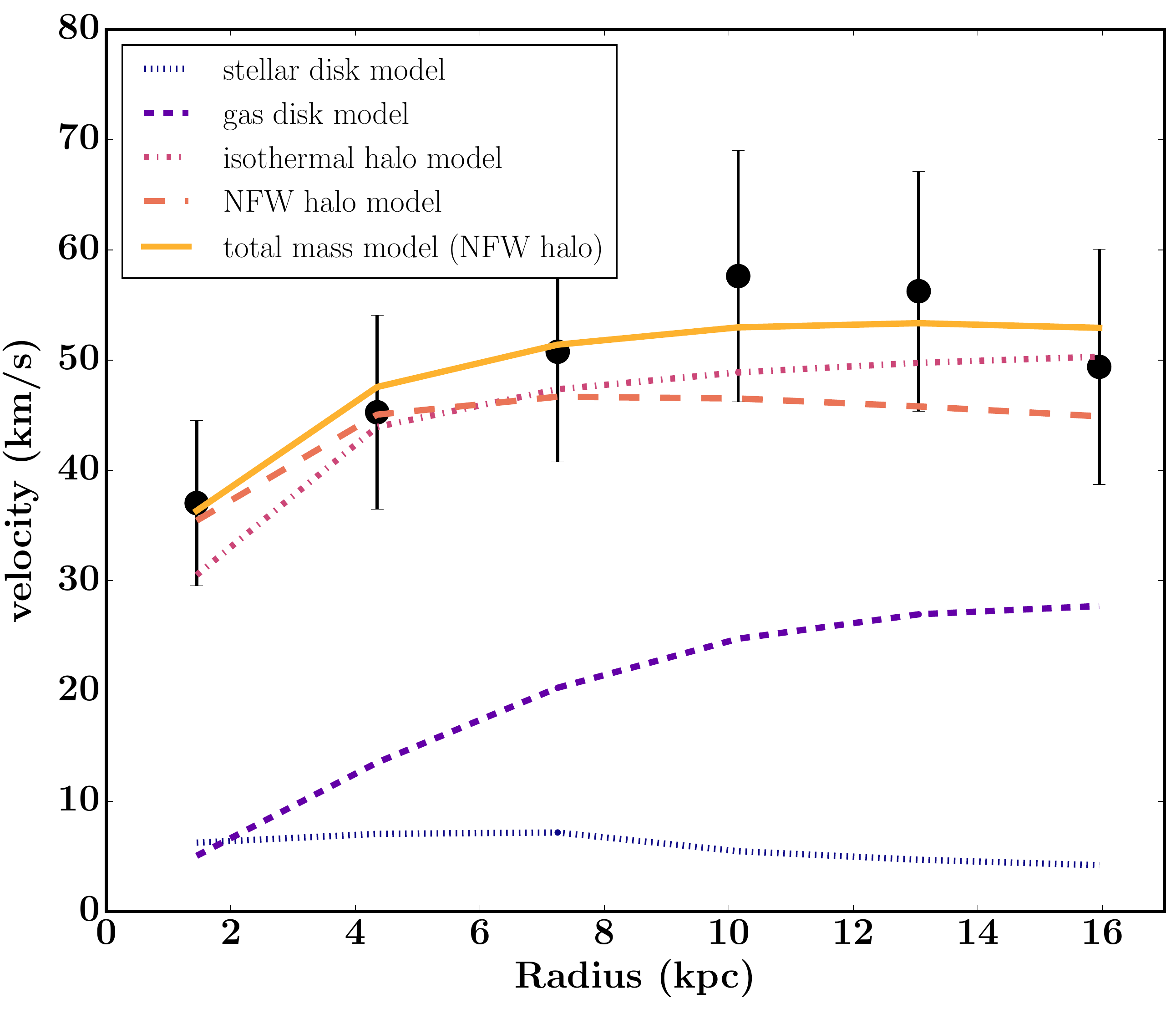}
\includegraphics[width=0.53\textwidth]{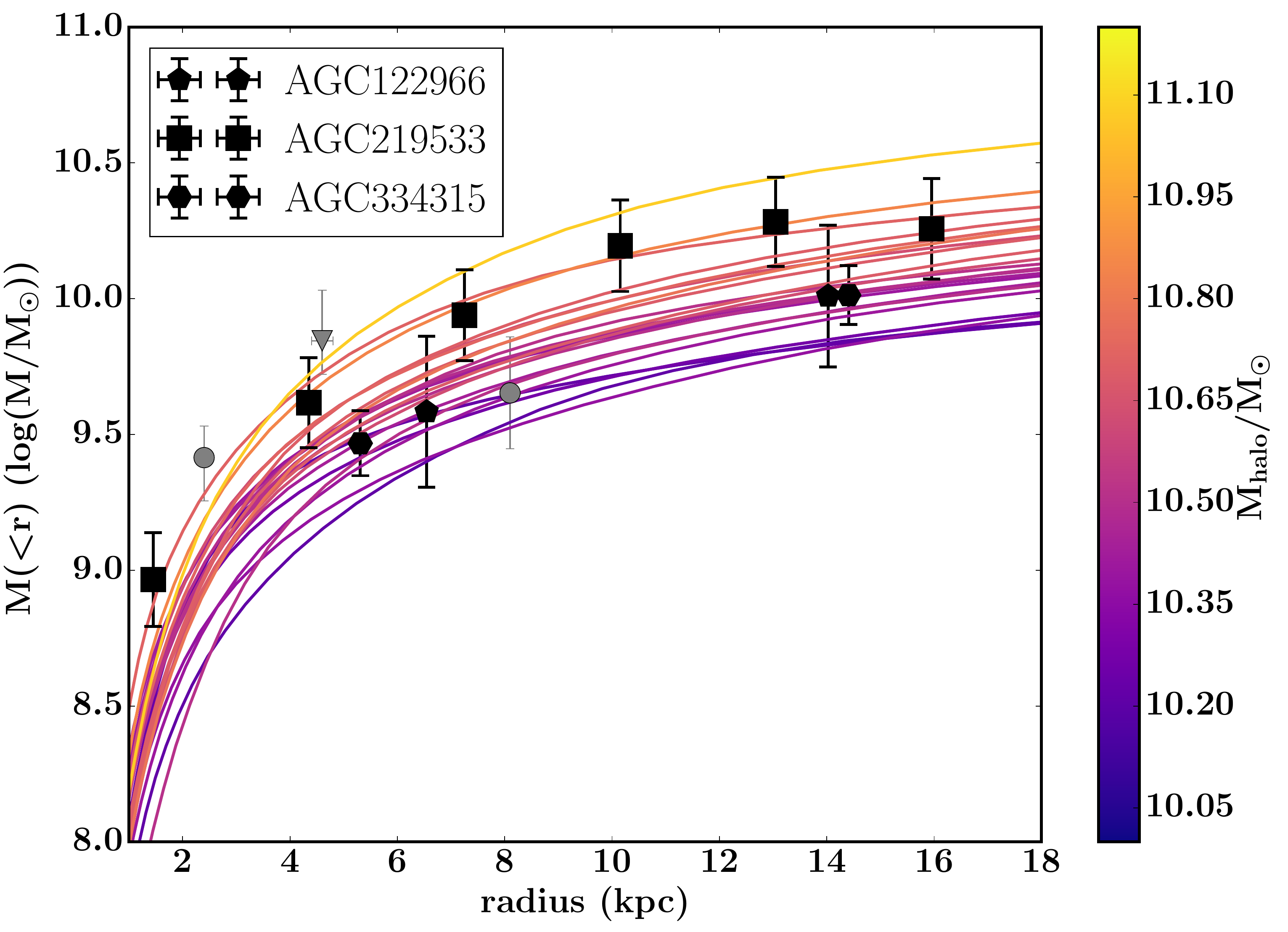}\\
\caption{Left: Rotation curve decomposition for AGC~219533. The modeled \hi\ disk accounts for a large portion of the assumed rotation, and suggests a halo with a low concentration parameter. Right: Comparison of calculated mass profiles for the sources with resolved synthesis imaging (black squares, pentagons, and hexagons), versus the predicted profiles from \cite{dicintio16a} based on NIHAO simulations \citep{wang15a}. The dynamical mass estimates from \cite{beasley16a} and \cite{vandokkum16a} are marked by grey circles and a grey triangle respectively. The measured and predicted profiles show good agreement, suggesting halo masses below 10$^{11}$\msun.  
\label{dmhalo}
}
\end{figure*}

The properties of the host dark matter halos of UDGs are poorly understood. As discussed in section \ref{intro}, the strongest constraints come from dynamical studies of globular clusters and from scaling relations with globular cluster counts, but are limited by the spatial extent of the globular cluster distribution. Here we attempt to constrain the properties of the dark matter halos of the HUDS. While the three sources observed with synthesis telescopes are only resolved with 3-6 beams, the large extent of the sources still allows us to significantly constrain the halos out to $\sim$20~kpc.

\subsubsection{The Halo Profile and Mass}
\label{discussion.dm.mass}

The right hand panel of Figure \ref{dmhalo} shows the dynamical mass of the HUDS as inferred from the \hi\ rotation curves estimated in section \ref{results.hi.rot} (black squares, pentagons, and hexagons), compared with the dynamical mass estimates from globular cluster spectroscopy for UDGs from \cite{vandokkum16a} (DF~44; grey triangle) and \cite{beasley16a} (VCC~1287; grey circles), and predicted models from \cite{dicintio16a} (colored lines). The uncertainties in geometry dominate the uncertainties in estimating the rotation velocity. However, even accounting for these, the \hi\ data provides significant constraints on the dark matter masses. All three sources have measured dynamical masses consistent with or slightly larger than the measurements from \cite{beasley16a} of VCC~1287, and slightly smaller than the measurement from \cite{vandokkum16a}. Specifically, \cite{beasley16a} find a
dynamical mass of 4.5$\pm$2.8$\times$10$^9$\msun\ within 8.1~kpc, while the HUDS give values ranging from 5-10$\times$10$^9$\msun\ within a similar radius (and with similar errors - see Table \ref{synthesistable}).
These dynamical estimates yield dynamical to stellar mass ratios for these sources consistent with those reported in \cite{beasley16a}, though the error bars are large. The dynamical to baryonic mass ratios are significantly smaller, however, since the \hi\ mass is large compared with the stellar mass.

The HUDS also match reasonably well with the predicted profiles from \cite{dicintio16a}, 
though for the best resolved source AGC~219533, the measured rotation curve appears somewhat flatter than those predicted. We estimate a halo mass from the best fitting profiles from \cite{dicintio16a} of $\sim$10$^{10.7}$\msun for AGC~219533, and somewhat smaller masses of 10$^{10.4}$ and 10$^{10.3}$\msun for AGC~334315 and 122966 respectively. 
However, the extrapolation from dynamical mass to total halo mass necessarily relies on the type of model fit. We note that abundance matching (using the data from \citealp{papastergis12a}) implies that galaxies with the baryonic masses of these sources should live in halos with log M$_{halo}$/\msun\ $\gtrsim$11.1. However, this estimate seems unreasonably large given the dark matter mass estimated within r=20~kpc.

 As an instructive exercise, we attempt to model the rotation curve as composed of a gaseous disk, stellar disk, and dark matter halo using the GIPSY task ROTMAS for AGC~219533. The left hand panel of Figure \ref{dmhalo} shows the best fitting model stellar and gas disk contributions to the rotation, assuming the mass surface density distributions shown in Figure \ref{surfacedensity}, multiplied by 1.3 to account for the presence of helium (and assuming an infinitely thin disk, an assumption that has little effect on the analysis given the size of the errors). The remaining rotation is modeled as the result of either an pseudo-isothermal or Navarro--Frenk--White (NFW; \citealp{navarro97a}) halo (shown as dash-dotted and long dashed lines respectively), such that $V_{obs}^2=V_{\rm gas}^2+V_*^2+V_{\rm DM}^2$.  

We also note that while low surface brightness galaxies usually exhibit steadily rising rotation curves that are not well fit by NFW profiles (e.g., \citealp{mcgaugh98a}), AGC~219533 appears to flatten out, potentially suggesting that is is more consistent with the NFW profile, similar to the extreme gas-rich, high spin parameter, low surface brightness galaxy UGC~12506 \citep{hallenbeck14a}. However, while these results are suggestive, the limited resolution of our current observations cautions against over interpreting these results.



\subsubsection{The Halo Spin Parameter}
\label{discussion.dm.spin}

While the analysis in section \ref{discussion.dm.mass} indicates that HUDS are likely to occupy dwarf halos, it does not explain the mechanism for the extended radii. 
Here we attempt to estimate the spin parameters of the dark matter halos of HUDS, to test the prediction that ``ultra-diffuse" sources are spatially extended due to large halo spin parameters \citep{amorisco16a}.

The spin parameter is a dimensionless quantity that describes the angular momentum in the halo:
$$\lambda = \frac{J*|E|^{1/2}}{G*M^{5/2}}$$
where J is the halo angular momentum, E is the energy, and M is the halo mass. The halo spin parameter is difficult to constrain observationally, since almost any $\lambda$ can fit a halo with given parameters from rotation curve fitting. Instead, we employ the common practice of simplifying the calculations by assuming that the dark and baryonic matter are coupled such that their angular momentum per unit mass ($j$ and $j_b$ respectively) are equal (e.g., \citealp{mo98a}). 
Thus we are technically calculating the modified spin parameter $\lambda' = j_b/j \times \lambda$ (henceforth, we will drop the prime). 

\begin{figure*}[]
\centering
\includegraphics[width=0.45\textwidth]{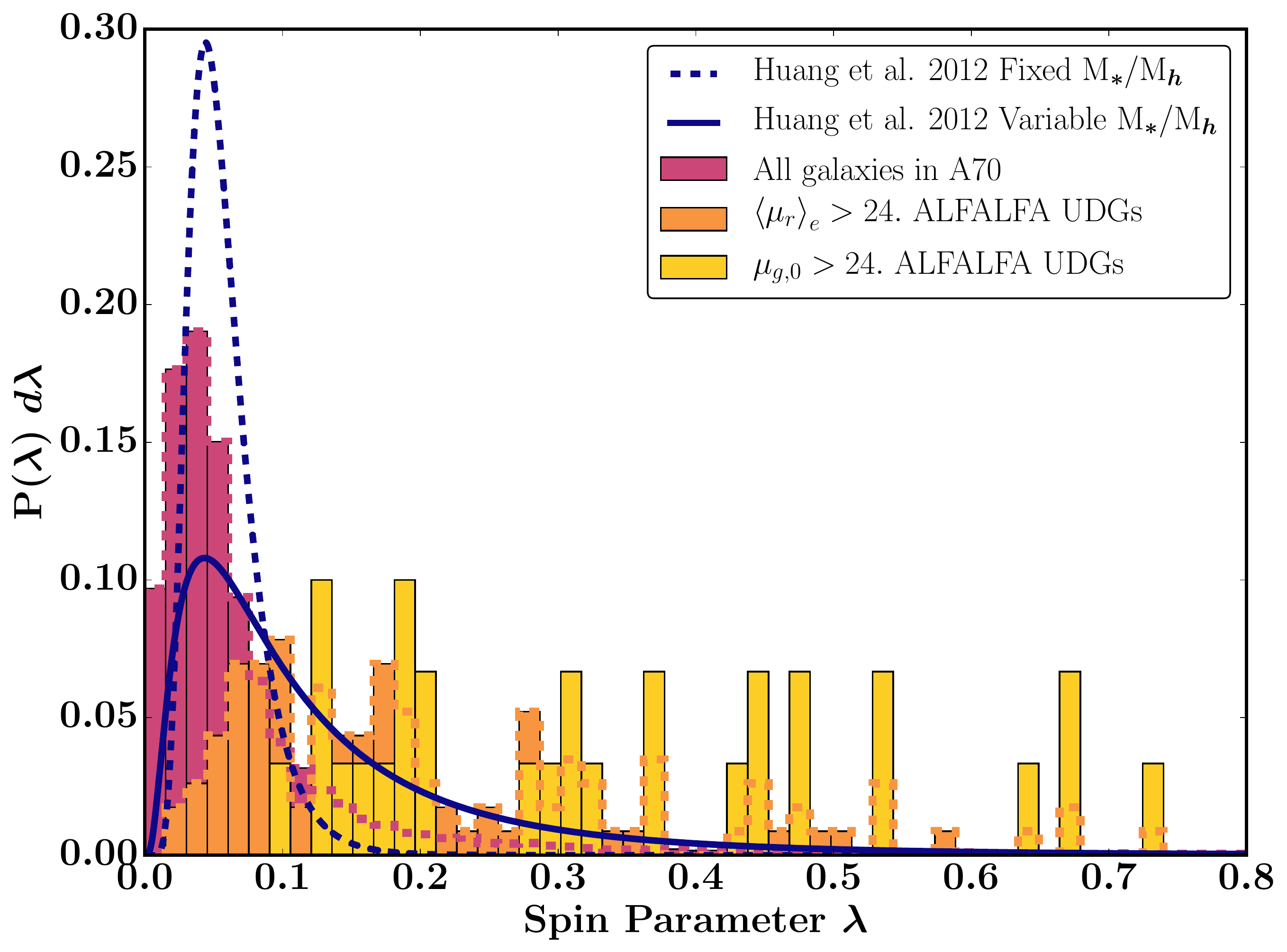}
\includegraphics[width=0.45\textwidth]{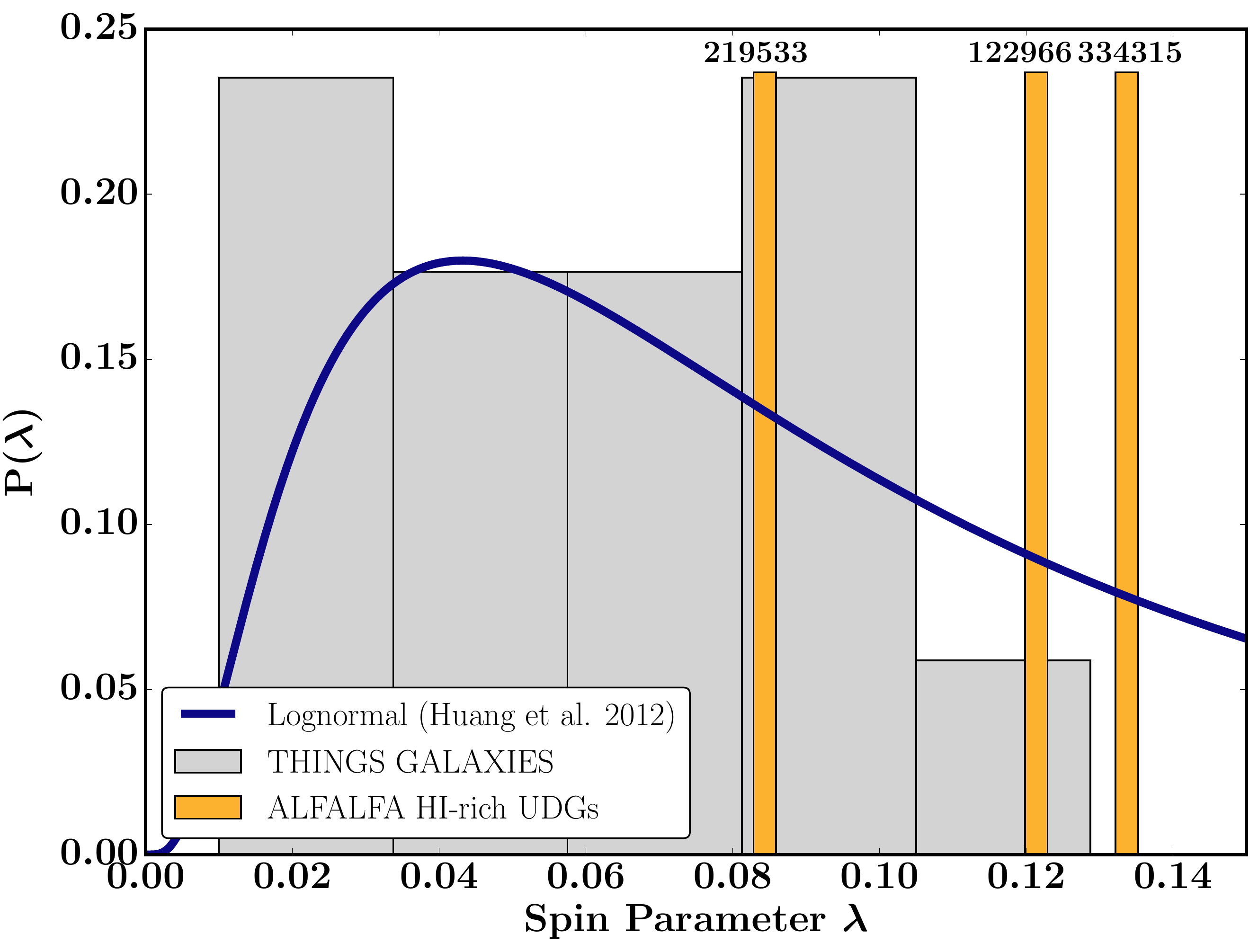}\\
\caption{Left: Spin parameter distribution of the HUDS-B and HUDS-R samples compared with the distribution for all ALFALFA sources. Spin parameters are calculated assuming an inclination of 45$^{\circ}$ following \cite{huang12b}. The blue dashed curve shows the distribution from \cite{huang12b} for a volume limited \hi\ selected sample, and the blue solid curve shows the result from \cite{huang12b} under the assumption of a variable halo mass fraction. Right: Comparison of the spin parameters for the three resolved HUDS to the THINGS sample, calculated using the method from \cite{hallenbeck14a}. The solid blue curve is the same as in the left hand panel. These results may suggest that these isolated, HUDS preferentially reside in high spin parameter halos. 
\label{spinparameter}
}
\end{figure*}

Under this assumption, we approach the calculation of $\lambda$ two different ways. We first follow \cite{huang12b}, measuring the exponential disk scale length $R_d$ from SDSS and the rotation velocity $V_{rot}$ from the ALFALFA \hi\ line width ($V_{rot} = W_{50}/2/sin(i)$), and adopting the $\lambda$ estimator from \cite{hernandez07a}:
$$\lambda=21.8 \frac{R_d [kpc]}{V_{rot}[km/s]^{3/2}}$$
This estimator further assumes self-gravitating, virialized, isothermal dark matter halos that dominate the galaxy's potential energy, flat rotation curves, and, importantly, a constant disk mass fraction M$_*$/M$_{\rm total}$ of 0.04. We assume that all sources are inclined at 45$^{\circ}$, as discussed in section \ref{data.inc}. 

The left panel of Figure \ref{spinparameter} shows the results of this analysis. The distribution for all galaxies in the ALFALFA 70\% sample is shown in dark pink. The distribution for the HUDS-R and HUDS-B samples are show in yellow and orange respectively, and appear to be significantly elevated relative to the rest of the ALFALFA distribution, i.e., the the radii of these sources are large given the rotation velocities of their disks. A K-S test estimates a probability that they are drawn from the same distribution as 10$^{-18}$ and 10$^{-34}$. 

To better understand the potential impact of selection effects and our other assumptions, the left hand panel also shows the lognormal fit to the spin parameter distribution of \hi\ selected sources derived by \cite{huang12b} through similar analysis (blue dashed curve). 
Like the spin parameters derived by \cite{huang12b}, the ALFALFA 70\% spin parameters follow a lognormal distribution, but have a lower mean and wider dispersion, demonstrating the impact of source selection.  The distribution shown in pink includes the entire ALFALFA sample, whereas the blue dashed curve is restricted to a volume limited sample from the ALFALFA 40\% catalog, and only includes sources over the absolute magnitude range -20~$>$~M$_r>$~-23~mag. The difference between the distributions is also in part due to the assumption of constant inclination. 

Further, the left hand panel also shows the lognormal distribution derived by \cite{huang12b} under the assumption the sources do not have a constant halo mass fraction, but instead have mass fractions derived from abundance matching (dark blue solid curve). As discussed in \cite{huang12b}, allowing the mass fraction to vary can have a large impact on the distribution. 
Still, the calculated spin parameters for the HUDS are large compared to the overall sample in all three cases, though we note that if the HUDS reside in large halos for their stellar masses as predicted through abundance matching, their estimated spin parameters would be significantly lower.
 
The trend to high spin parameters is perhaps not entirely unexpected given our selection of extended sources,
and the observation that they have high gas fractions relative to the ALFALFA sample (section \ref{results.hi}). Indeed, \cite{huang12b} showed that sources with high gas fractions tend to have high spin parameters for a given stellar mass, and \cite{obreschkow16a} suggest that gas fraction depends on a global stability parameter which scales linearly with the angular momentum of the disk. Thus, the high estimated spin parameters make sense in light of the sample's observed gas fractions.

For resolved sources we can do a somewhat more detailed estimate of the modified halo spin parameter.  We follow the procedure detailed in \cite{hallenbeck14a}, which,  in brief, estimates the angular momentum of the halo by summing the product of the \hi\ disk mass, velocity, and radius at each point on the rotation curve, and assuming the angular momentum of the halo scales with that of the baryons. The energy is calculated from the maximum velocity from the isothermal halo fit, and the halo mass is estimated from abundance matching. We note that, as discussed in section \ref{discussion.dm.mass}, abundance matching poorly estimates the total halo mass for these sources. Thus we also estimate spin parameters assuming the halo masses derived from comparison to the models from \cite{dicintio16a}, and discuss the effect on the results below. 

The right hand panel of Figure \ref{spinparameter} shows the distribution of spin parameters for resolved THINGS galaxies computed by \cite{hallenbeck14a}, which approximately follow the lognormal distribution for the 40\% ALFALFA sample (also plotted in the left panel) from \cite{huang12b}.  The three HUDS with synthesis observations, assuming the halo masses from \cite{dicintio16a}, are overplotted as yellow bars. Though this analysis uses resolved rotation curve fitting rather than SDSS radii and ALFALFA linewidths, the 3 sources again appear to have higher spin parameters than most of the THINGS galaxies. 

However, it is important to note that if one instead assumes the halo masses from abundance matching, which are larger by a factor of $\sim$4, the spin parameters are reduced by the same factor, and fall squarely within the THINGS distribution. Also, for 2 of the 3 sources, the values estimated from our resolved analysis are significantly lower than those estimated from the velocity width and optical radius. This is likely due to the low physical resolution relative to the sources from \cite{hallenbeck14a}, which tends to suppress the j value, and the comparatively large r$_{\rm optical}$/r$_{\rm HI}$, which increases the spin parameter measurement in the unresolved method (which relies on optical radii) relative to the resolved method (which relies on \hi\ radii measurements). With only 3 sources, we hesitate to read more into this.

With these caveats in mind, the results in this section still suggest that if the assumptions used to calculate the spin parameters are valid and the halo masses are indeed more consistent with dwarfs, then HUDS may reside in high spin parameter halos.

\subsection{The Nature of Isolated HI-bearing Ultra-Diffuse Sources}

While further observations will be necessary to fully understand the connection between HUDS and other UDGs, the observations presented here 
are consistent with the predicted population of reasonably isolated UDGs with large gas fractions from \cite{dicintio16a}. 
More, these data begin to paint a picture of isolated HUDS as extremes in the dwarf galaxy population. Their dynamical mass estimates suggest that at least the 3 resolved HUDS are inconsistent with being failed L$_{\star}$ galaxies. While this result necessarily relies on assumptions about the disk thickness, and the effects of the beam, in general,  these systematic uncertainties would tend to reduce the estimated halo mass. The effect of a thick disk would be to underestimate the inclination, thus overestimating our rotational velocities. Additionally, though beam smearing can tend to underestimate velocity gradients and overestimate velocity width, our envelope fitting technique (section \ref{results.hi.rot}) functions as an effective upper limit on the velocity gradient. 

Further, while the large \hi\ masses and correspondingly large radii are consistent with a range of halos, the large gas fractions and the star forming characteristics of the HUDS seem more consistent with sources of their stellar mass than their \hi\ mass. Though their low SFE is what we might expect if these sources were failed L$_{\star}$ galaxies, it also consistent with low density dwarfs or sources that have recently experienced gas accretion.

These results, in turn, support potential scenarios that connect HUDS to other UDGs in dwarf halos. One potential scenario is one where isolated gas-bearing ``ultra-diffuse" sources continue to inefficiently form stars until they fall into clusters or groups and have their gas stripped, quenching star formation. In time, the blue colors would fade and the clumpy morphologies would disappear. Indeed, \cite{roman16b} recently estimate that 6 ``progenitor" UDG sources on the edges of groups (with properties somewhat similar to the HUDS-B sample), might fade $\sim$1.5~mag~arcsec$^{-2}$ if they evolve passively for 6~Gyr. 

Then again, it is also possible that HUDS are an independent population. In this scenario, we may be observing them at a particularly interesting period of gas accretion before they are transformed by significant star formation. Thus, it would undergo a significant increase in surface brightness as it evolves. Detailed, high resolution study of the gas dynamics of these sources will be necessary to explore this possibility further.

Yet another possibility is that the HUDS are not a uniform physical population, and instead result from multiple independent mechanisms. During the review process for this work, \cite{trujillo17a} and \cite{papastergis17a} have also reported the detection of \hi-bearing ``ultra-diffuse" sources, but their connection to the HUDS presented here is not yet clear. The source from \cite{trujillo17a}, UGC~2162 appears smaller and brighter than the sources in this sample - perhaps suggesting some connection to the smaller SHIELD galaxies (e.g., \citealp{cannon11a}; \citealp{mcquinn15a}; \citealp{teich16a}), and \cite{papastergis17a} suggest the possibility of at least two populations of isolated UDGs, pointing to the need for significant future work in this field.

\section{Conclusions}
\label{conclusions}

Here we investigate the properties of isolated, very low surface brightness, ``ultra-diffuse" galaxies detected in the ALFALFA survey, and present follow up observations of three of these extreme sources. The main conclusions of this paper are:
\begin{enumerate}
\itemsep0em
\item There exists a substantial population of \hi-bearing ultra-diffuse sources (HUDS) with similar surface brightnesses, masses, and radii to recently reported ``ultra-diffuse galaxies." We select samples of sources from ALFALFA to match optical selection criteria of ``ultra-diffuse" galaxies and find 30-115 HUDS (depending on where we define the surface brightness cut) in the $\sim$5000 isolated ALFALFA galaxies with 25$<{\rm Dist}<$120~Mpc.
\item The HUDS are significantly bluer and have more irregular morphologies than their non-isolated counterparts. They appear to be forming stars at typical rates of \hi-selected galaxies for their stellar mass.
\item The HUDS are \hi-rich for their stellar mass; the sources have elevated M$_{HI}$/M$_{*}$ ratios. Thus, these galaxies have very low star formation efficiencies, with gas consumption timescales longer than a Hubble time.
\item  The three resolved HUDS have large \hi\ disks, that extend well beyond their diffuse stellar counterparts, and are similar to the \hi\ radii measured for L$_{\star}$ spiral galaxies. The \hi\ appears to be relatively ``diffuse" and low column density, but it is not ``ultra-diffuse"; its extent is only slightly larger than predicted by \hi\ mass-radius scaling relations.
\item HUDS have relatively narrow velocity widths compared with the rest of the ALFALFA sample, even when correcting for inclination and mass selection effects. This, coupled with rough dynamical modeling of the three resolved HUDS, suggests that though their \hi\ and optical diameters are similar to L$_{\star}$ galaxies, they have dynamical masses consistent with the smaller dwarf halos expected given their stellar mass. However, we note that given the poor resolution of the current observations, it is not possible to disentangle possible effects of gas infall and non-standard disk geometry.
\item The combination of large radii and low rotation velocities suggests, under the assumption that the angular momentum of the disk traces the angular momentum of the halo, that these HUDS reside in high spin parameter halos, potentially implying that the high angular momentum of the halos is responsible for their ``ultra-diffuse" nature.
\end{enumerate}
\vskip -5pt
Together, these observations suggest that these isolated HUDS are gas-rich, low density, extended dwarfs, in unusual lower mass halos. Therefore they may be related to gas poor, non-isolated UDGs with similar halo masses. However, further observations and modeling will be required to understand the nature of that connection, and their place in the evolutionary history of very low surface brightness galaxies.

\acknowledgments
{\bf Acknowledgments}.  The authors acknowledge the work of the entire ALFALFA
collaboration in observing, flagging, and extracting sources. 
The authors would like to thank Arianna Di Cintio and the NIHAO collaboration for sharing their simulated mass profiles,
and the anonymous referee for useful suggestions that improved the quality of the paper. 
LL would like to thank Michael G. Jones for several useful discussions.
The ALFALFA team at Cornell is supported by NSF grants AST-0607007 
and AST-1107390 to RG and MPH and by grants from the Brinson Foundation. 
SJ acknowledges support from the Australian Research Council’s Discovery Project funding scheme (DP150101734).
JMC is supported by NSF grant AST-1211683. 
EAKA is supported by TOP1EW.14.105, which is financed by the Netherlands Organisation for Scientific Research (NWO).
DBN is supported by the Office of the Vice President for Research, Universidad de los Andes.
W.F.J. and K.L.R. are supported by NSF grant AST-1615483.

This work is based in part on observations made with the VLA, Arecibo Observatory, and WSRT.
The VLA is a facility of the National Radio Astronomy Observatory (NRAO). NRAO is a facility of the National Science Foundation operated under cooperative agreement by Associated Universities, Inc.  The Arecibo Observatory is operated by SRI International under a cooperative agreement with the National Science Foundation (AST-1100968), and in alliance with Ana G. M\`endez-Universidad Metropolitana, and the Universities Space Research Association. 
The Westerbork Synthesis Radio Telescope is operated by the ASTRON (Netherlands Institute for Radio Astronomy) with support from NWO.

This research used data from the Sloan Digital Sky Survey, funded by
the Alfred P. Sloan Foundation, the participating institutions, the
National Science Foundation, the U.S. Department of Energy, the
National Aeronautics and Space Administration, the Japanese
Monbukagakusho, the Max Planck Society, and the Higher Education
Funding Council for England.

\facilities{Arecibo, WSRT, WIYN}
\software{IDL, Python, Astropy, APLpy, Miriad}

\bibliography{mybib}
\bibliographystyle{apj}

\appendix
\section{Comparison of 1D Profiles derived from SDSS and WIYN Images}

\begin{figure}[]
\centering
\includegraphics[width=0.49\textwidth]{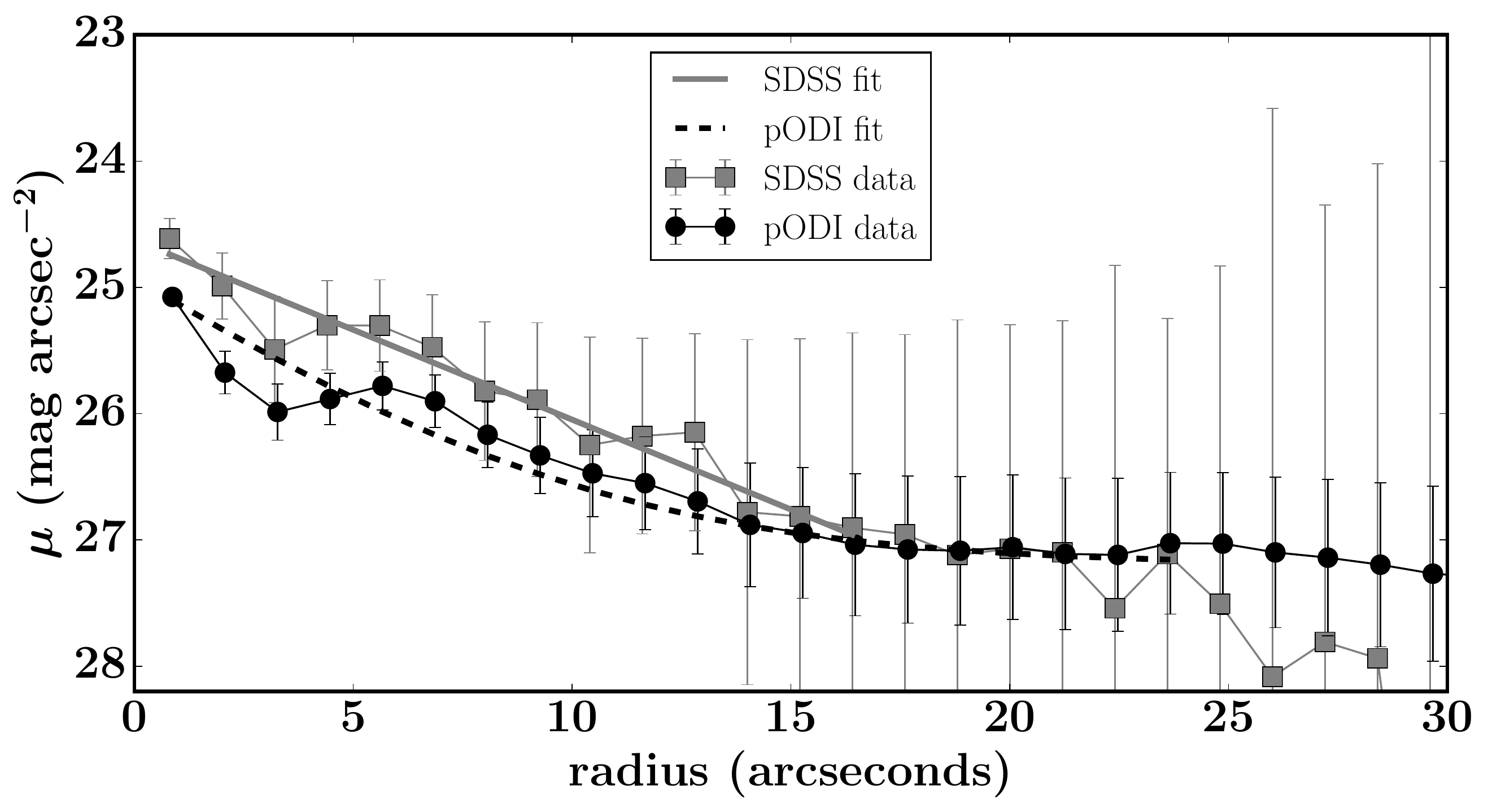}
\includegraphics[width=0.49\textwidth]{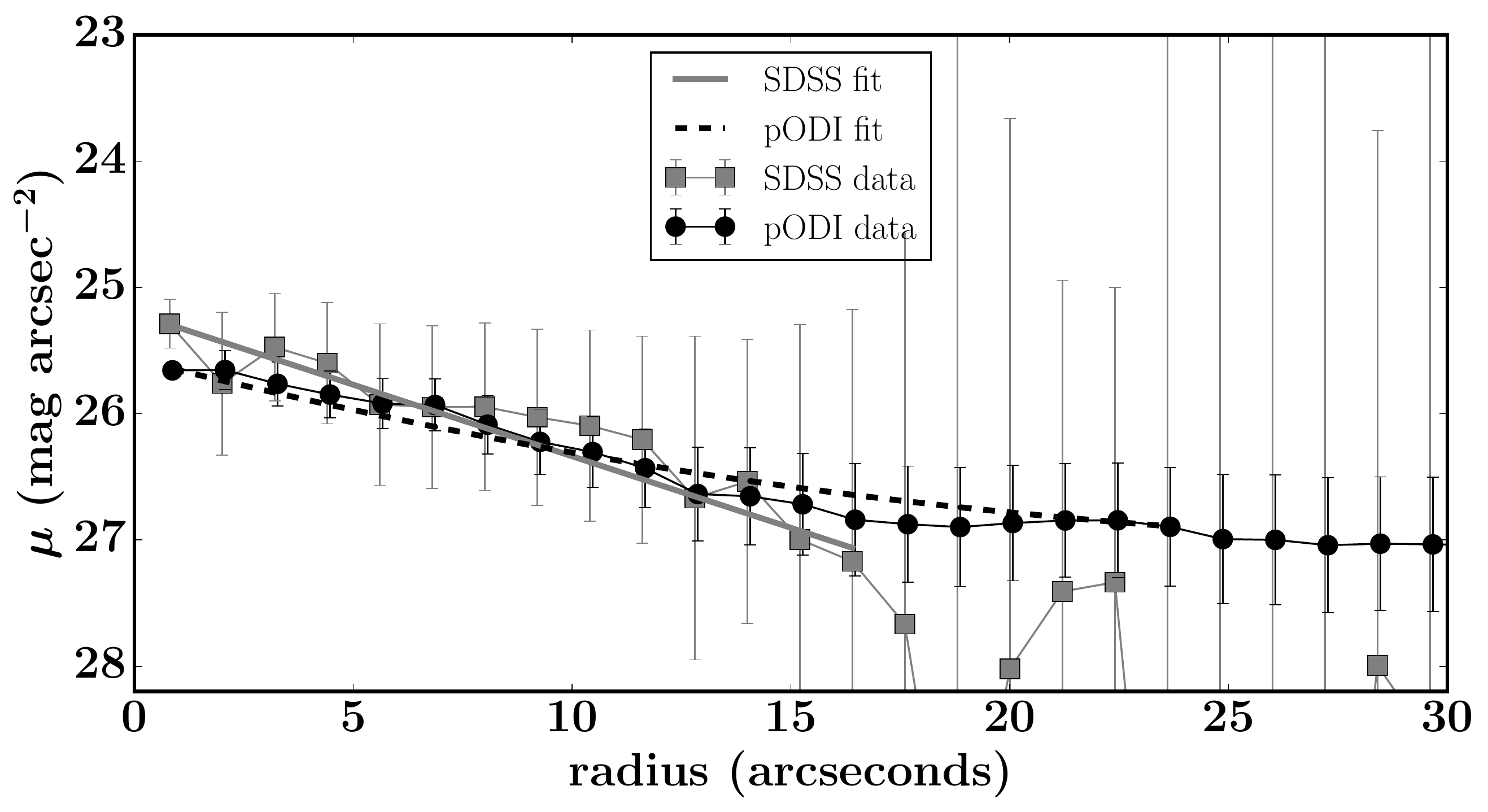}\\
\caption{Comparison of 1D profiles derived from circular apertures on SDSS and deep WIYN images for AGC~334315 (left) and AGC~122966 (right). These profiles demonstrate the rough reliability of the SDSS data for sample selection, and the need for deeper data to obtain detailed structural information about these very low surface brightness, irregular sources.  
\label{fig.profcompare}
}
\end{figure}

Since the sources in the HUDS samples are only barely detected in SDSS, here we 
explore the reliability of the SDSS measurements by comparing our measured 
profiles to deeper WIYN imaging.
While it is clearly true that these galaxies are very low surface brightness and 
very extended from visual inspection, this comparison provides a rough test of 
our quantitative estimates of surface brightness and radius
for purposes of sample selection.

Specifically, for both sources with deep WIYN imaging we apply our simple 1D 
fitting procedure using identical apertures to both the SDSS and WIYN images.
Figure \ref{fig.profcompare} shows the resulting profiles and 1D fits for both
sources. 

The profiles show g-band data, and cut off the profile fits when the signal drops to 0. For AGC~122966 the fits are almost entirely consistent within the errors. For AGC~334315 the SDSS data is systematically offset to brighter values than the ODI data. This is because AGC~334315 is very close to a bright star, which significantly affects its flux measurement, and creates significant uncertainty in measurements of the local background. This, combined with AGC~334315's relatively low surface brightness makes it a ``worst case" scenario, but even in this case the measurements are reasonably consistent. For AGC~334315 (122966) we find a central surface brightness of 24.6+/-0.2 (25.2+/-0.2) mag~arcsec$^{-2}$ using the SDSS image, and 24.8+/-0.1 (25.5+/-0.1) mag~arcsec$^{-2}$ using the WIYN image. We derive effective radii if 7.6+/-2.1 (9.6+/-4.2)~kpc from SDSS, and 4.6+/-0.6 (9.9+/-2.1)~kpc from WIYN. This agreement seems especially good given the irregularity morphology of the stellar disks and uncertainties in the background subtraction.

Still, we emphasize that the main point of this paper is not an in depth study of the detailed structural parameters of these sources - the SDSS data are insufficient for this purpose. Thus, these profiles are not intended to provide detailed structural information, but rather to show the rough 
reliability of the data for purposes of sample selection, and also to qualitatively demonstrate our
uncertainties.

\end{document}